\documentclass[widetext,showpacs,preprintnumbers,superscriptaddress,nofootinbib]{revtex4}

\usepackage[dvipdfm]{graphicx}
\usepackage{amssymb}
\usepackage{amsmath}
\usepackage{dcolumn}
\usepackage{bm}
\usepackage{natbib}
\usepackage{multirow}
\usepackage{subfigure}

\begin{document}

\preprint{ICRR-Report-632-2012-21}

\title{Evolution and thermalization of dark matter axions in the condensed regime}

\author{Ken'ichi Saikawa}
\email{saikawa@icrr.u-tokyo.ac.jp}
\affiliation{Institute for Cosmic Ray Research, The University of Tokyo, 
5-1-5 Kashiwa-no-ha, Kashiwa City, Chiba 277-8582, Japan}
\author{Masahide Yamaguchi}
\email{gucci@phys.titech.ac.jp}
\affiliation{Department of Physics, Tokyo Institute of Technology,
2-12-1 Ookayama, Meguro-ku, Tokyo 152-8551, Japan}
 
\date{\today}

\begin{abstract}
We discuss the possibility that dark matter axions form a Bose-Einstein
condensate (BEC) due to the gravitational self interactions. The
formation of BEC occurs in the condensed regime, where the transition
rate between different momentum states is large compared to the energy
exchanged in the transition. The time evolution of the quantum
state occupation number of axions in the condensed regime is derived
based on the in-in formalism. We recover the expression for the
thermalization rate due to self interaction of the axion field, which
was obtained in the other literature. It is also found that the leading
order contributions for interactions between axions and other species
vanish, which implies that the axion BEC does not give any significant
modifications on standard cosmological parameters.
\end{abstract}


\maketitle


\section{\label{sec1}Introduction}

Identifying the origin of the dark matter of the Universe is one of the
priorities of modern high-energy physics and astronomy.  So far, many
particle physics models of the dark matter have been proposed (see
e.g.~\cite{Bertone:2004pz} for reviews), and well-motivated candidates
are so-called weakly interacting massive particles (WIMPs) and
axions. Both of them possess suitable properties for the dark matter in
that they are nonbaryonic, cold, and collisionless. However, the nature
of cosmological behavior is completely different between WIMPs and
axions. WIMPs are produced from a primordial soup of radiations, and their
population is fixed when they decouple from the thermal plasma.  We can
interpret them as a collection of classical particles, whose velocity
dispersion is determined by the decoupling temperature.  On the other
hand, axions are produced nonthermally, having a small velocity
dispersion compared with the temperature of the thermal plasma at the
epoch when they are produced. Furthermore, they have a huge occupation
number in the phase space since they are bosons.  Because of these
properties, we interpret them as a classical field rather than
individual particles. This classical field of axions coherently
oscillating in the field space behaves as a cold matter component of the
Universe~\cite{Preskill:1982cy}.

These curious properties of dark matter axions motivate the possibility
that axions form a Bose-Einstein condensate (BEC)~\cite{Sikivie:2009qn}.
Indeed, the energy dispersion of axions at the production time
$\delta\omega\sim {\cal O}(10^{-13})\mathrm{eV}$ is much smaller
than the critical temperature for BEC,
$T_c=(\pi^2n/\zeta(3))^{1/3}\sim {\cal O}(100)\mathrm{GeV}$,
where $n$ is the number density of axions.\footnote{Here, the expression for the critical temperature $T_c=(\pi^2n/\zeta(3))^{1/3}$ used in~\cite{Sikivie:2009qn}
is applied for relativistic particles, which might not be appropriate for cold axions
because they are non-relativistic. Even though, the condition $\delta\omega\ll T_c$
is also satisfied if we use the expression for non-relativistic particles,
$T_c=(2\pi/m)(n/\zeta(3/2))^{2/3}\sim{\cal O}(10^{22})\mathrm{GeV}$
where $m$ is the mass of the axion.}
However, this argument
relies on the following assumptions. First, the particles must be
bosons. Second, their number must be conserved.  Third, they should have
huge phase space density. Finally, they must be in thermal equilibrium.
The first three conditions are obviously satisfied for dark matter
axions, but the final one, whether axions thermalize or not, is a
nontrivial issue.

The cosmic thermalization of axions is extensively discussed in
Ref.~\cite{Erken:2011dz}. Here, the thermalization means that the system
relaxes into a state with the highest entropy by exchanging energies and
momenta between particle states. In order to investigate such a process,
some statistical mechanical treatments are required.  The usual analysis
using the Boltzmann equation cannot be applied to this problem, since
highly degenerate axions are essentially fields in the classical limit.
Such a system of axion fields does not match the assumption of the
Boltzmann analysis, where the system is considered to be a collection of
point particles. This situation defines a peculiar regime of the many
body system, called the condensed regime, which should be distinguished
from the particle kinetic regime where the Boltzmann analysis can be
applied. In the condensed regime, the energy transfer of the scattering
process is small compared with the scattering rate, because the
interaction occurs between highly degenerate states. Thermalization in
the condensed regime is not well understood in comparison with that of
the particle kinetic regime.

In Ref.~\cite{Erken:2011dz}, the thermalization process in the condensed
regime is analyzed by describing the axion field as a set of
quantum-mechanical oscillators (i.e. quantum operators) and deriving the
evolution equation of each oscillator.  However, in the formalism
of~\cite{Erken:2011dz} the thermalization rate is estimated by comparing
the ``orders of magnitude'' of quantum operators, and the occurrence of
the thermalization is confirmed only by computing the quantum-mechanical
averages of the occupation number of each oscillator numerically, which
is realized for toy models with a small number of oscillators and
particles.  It is impossible to realize the actual system with a huge
number of axions using the numerical scheme described
in~\cite{Erken:2011dz}.

In this paper, we revisit the issue of the axion thermalization. The
purpose of this work is to develop a robust tool to describe the
relaxation process of highly degenerate bosonic fields. Instead of using
the approach of Ref.~\cite{Erken:2011dz}, we compute the expectation
value (i.e. the quantum-mechanical average) of the occupation number of
axions, and solve its time evolution, which informs us of the
change of the distribution function of axions. The computation is
executed by using the technology which was originally introduced by
Schwinger {\it et al}.~\cite{Schwinger:1960qe}, called the ``in-in''
formalism. This formalism enables us to calculate the time evolution of
the expectation value of quantum operator in the systematic way.
Furthermore, it can treat a state to which particlelike interpretation
is not applied, if we use an appropriate representation for a state at
the initial time. In our case, a coherent state is used to describe a
state in the condensed regime. Using this formalism, we estimate the
thermalization rate as the inverse of the time scale in which the
expectation value of the occupation number changes its value.  Whole
things can be described in the analytic way, and it is not necessary to
use numerical simulations.

The discussion on thermalization of axions and formation of a BEC is not
only the theoretical issue, but has a relevance to observations.  There
are several observational evidences indicating that the phase space
structure of galactic halos is consistent with the ``caustic ring
model''~\cite{Duffy:2008dk}. In this model, the high density surfaces
(caustics) in the phase-space distribution of the dark matter particles
become ringlike configurations when dark matter particles fall into a
galactic potential with net overall rotation. Recently, it was pointed
out that this caustic ring model is predicted if dark matter axions form
a BEC~\cite{Sikivie:2010bq}, which might be regarded as an evidence of
the axion dark matter.  However, it was also suggested that an axion BEC
might enter into thermal contact with photons, and modify some
cosmological parameters from the standard
values~\cite{Erken:2011vv,Erken:2011dz}.  In particular, if axions and
photons reach thermal equilibrium, the photon temperature is cooled,
which predicts a smaller value of the baryon to photon ratio at the big
bang nucleosynthesis and a larger value of the effective number of
neutrino species $N_{\mathrm{eff}}$. The predicted value is
$N_{\mathrm{eff}}=6.77$, which is larger than the observed value
$N_{\mathrm{eff}}\simeq 3$-$5$~\cite{Komatsu:2010fb}.  This result seems
to be disapproval of axion BEC dark matter, but we find that such photon
cooling effects are fictitious. As will be shown later, axions do not
enter into thermal contact with photons, though they
develop toward a BEC.

The organization of this paper is as follows.  In Sec.~\ref{sec2}, we
introduce the method to evaluate the time evolution of the occupation
number of axions.  The expectation value of the quantum operator for the
occupation number is computed by using the perturbative expansion in
terms of the interaction Hamiltonian of the axion field. We perform the
calculation of the leading order terms in perturbation theory, while the
second order terms are evaluated in the Appendix. In
Sec.~\ref{sec3}, implications of the results of our analysis on
cosmology are discussed. We estimate the relaxation rate of
self-interacting degenerate axions and recover the formula for the
thermalization rate of axions, which was obtained in
Ref.~\cite{Erken:2011dz}. We will see that the thermalization rate
exceeds the expansion rate when the temperature of the Universe becomes
sufficiently low, at which axions begin to evolve toward a BEC.
Finally, the summary and
conclusions are given in Sec.~\ref{sec4}.

\section{\label{sec2}Axion field dynamics}

In this section, we develop the formalism to compute the time evolution
of quantum occupation number of the axion field.  Our interest is to
calculate the expectation value of a quantum operator $\hat{{\cal
N}}_{\bf p}(t)$ at a given time $t$, which represents the number of
axions occupying a quantum state labeled by three-momentum ${\bf p}$.
Such a problem can be dealt by using the in-in formalism (or
Schwinger-Keldysh formalism)~\cite{Schwinger:1960qe}. In cosmology, this
formalism was applied to calculate quantum contributions to cosmological
correlations~\cite{Weinberg:2005vy,Koyama:2010xj}. Following such
a formalism closely, in this work, we calculate the time evolution of
the occupation number in a systematic way by use of the perturbative
expansion.

The outline of this section is as follows. In Sec.~\ref{sec2-1}, we
review the formalism described in~\cite{Weinberg:2005vy,Koyama:2010xj}
and give the formula to calculate the expectation value of the quantum
operator. In Sec.~\ref{sec2-2}, the mode expansion of the field operator
is taken and the quantum occupation number is defined. In
Sec.~\ref{sec2-3}, we discuss how to represent the coherently
oscillating axion fields as quantum states.  Using the ingredients
obtained in Secs.~\ref{sec2-1}, \ref{sec2-2}, and \ref{sec2-3}, we
compute the time evolution of the occupation number due to the
self interaction of the axion field in Sec.~\ref{sec2-4}.  Finally,
interactions with other particles such as baryons and photons are
discussed in Sec.~\ref{sec2-5}.

\subsection{\label{sec2-1}The in-in formalism}

Let us consider the theory with a real scalar field (the axion field)
$\phi({\bf x},t)$ in the Minkowski background. The Lagrangian density is
given by
\begin{equation}
{\cal L} = -\frac{1}{2}\partial^{\mu}\phi\partial_{\mu}\phi - \frac{1}{2}m^2\phi^2 + {\cal L}_I, \label{eq2-1-1}
\end{equation}
where $m$ is the mass of the axion and ${\cal L}_I$ is the interaction
term which we specify later. The Hamiltonian of the system is given by
\begin{equation}
H[\phi(t),\pi(t)] = \int d^3x (\pi\dot{\phi}-{\cal L}) =  H_0[\phi(t),\pi(t)] + H_I[\phi(t),\pi(t)], \label{eq2-1-2}
\end{equation}
where $\pi({\bf x},t)=\dot{\phi}({\bf x},t)$ is the canonical
conjugate. Here, we decompose $H$ into its free and interaction terms,
\begin{align}
H_0[\phi(t),\pi(t)] &= \int d^3x \left[\frac{1}{2}\dot{\phi}^2+\frac{1}{2}|\nabla\phi|^2+\frac{1}{2}m^2\phi^2\right], \label{eq2-1-3}\\
H_I[\phi(t),\pi(t)] &=  - \int d^3 x{\cal L}_I. \label{eq2-1-4}
\end{align}

The quantum operators satisfy the canonical commutation relations,
\begin{align}
[\phi({\bf x},t),\pi({\bf y},t)] &= i\delta^{(3)}({\bf x-y}), \nonumber \\
[\phi({\bf x},t),\phi({\bf y},t)] &= [\pi({\bf x},t),\pi({\bf y},t)] = 0. \label{eq2-1-5}
\end{align}
Their time evolution is given by the Heisenberg equations,
\begin{align}
\dot{\phi}({\bf x},t) &= i\left[H[\phi(t),\pi(t)],\phi({\bf x},t)\right], \nonumber\\
\dot{\pi}({\bf x},t) &= i\left[H[\phi(t),\pi(t)],\pi({\bf x},t)\right]. \label{eq2-1-6}
\end{align}
These equations can be solved formally,
\begin{align}
\phi(t) = U^{-1}(t,t_0)\phi(t_0)U(t,t_0), \nonumber\\
\pi(t) = U^{-1}(t,t_0)\pi(t_0)U(t,t_0), \label{eq2-1-7}
\end{align}
for some fixed time $t_0$, where $U(t,t_0)$ is given by
\begin{equation}
\frac{d}{dt}U(t,t_0) = -iH[\phi(t),\pi(t)]U(t,t_0) \quad\mathrm{and}\quad U(t_0,t_0) =1. \label{eq2-1-8} 
\end{equation}

Now we move on to the interaction picture. Let us define interaction
picture fields $\phi^I$ and $\pi^I$ such that
\begin{align}
\dot{\phi}^I({\bf x},t) &= i\left[H_0[\phi^I(t),\pi^I(t)],\phi^I({\bf x},t)\right], \nonumber\\
\dot{\pi}^I({\bf x},t) &= i\left[H_0[\phi^I(t),\pi^I(t)],\pi^I({\bf x},t)\right], \label{eq2-1-9}
\end{align}
with $\phi^I(t_0) = \phi(t_0)$ and $\pi^I(t_0) = \pi(t_0)$.
Solutions of these equations are given by
\begin{align}
\phi^I(t) &= U_0^{-1}(t,t_0)\phi(t_0)U_0(t,t_0), \nonumber\\
\pi^I(t) &= U_0^{-1}(t,t_0)\pi(t_0)U_0(t,t_0), \label{eq2-1-10}
\end{align}
where $U_0(t,t_0)$ satisfies the following equation:
\begin{equation}
\frac{d}{dt}U_0(t,t_0) = -iH_0[\phi^I(t),\pi^I(t)]U(t,t_0) \quad\mathrm{and}\quad U_0(t_0,t_0) =1. \label{eq2-1-11} 
\end{equation}

By noting that
\begin{align}
H_0[\phi^I(t),\pi^I(t)] &= H_0[\phi(t_0),\pi(t_0)], \nonumber\\
H[\phi(t),\pi(t)] &= H[\phi(t_0),\pi(t_0)], \label{eq2-1-12}
\end{align}
Eqs.~(\ref{eq2-1-8}) and (\ref{eq2-1-11}) lead to
\begin{equation}
\frac{d}{dt}F(t,t_0) = -iH_I(t)F(t,t_0) \quad\mathrm{and}\quad F(t_0,t_0) = 1, \label{eq2-1-13}
\end{equation}
where
\begin{align}
F(t,t_0) &\equiv U^{-1}_0(t,t_0)U(t,t_0), \label{eq2-1-14}\\
H_I(t) &\equiv U_0^{-1}(t,t_0)H_I[\phi(t_0),\pi(t_0)]U_0(t,t_0) = H_I[\phi^I(t),\pi^I(t)]. \label{eq2-1-15}
\end{align}
The solution of Eq.~(\ref{eq2-1-13}) is given by
\begin{equation}
F(t,t_0) = T\exp\left(-i\int^t_{t_0} H_I(t)dt\right), \label{eq2-1-16}
\end{equation}
and also
\begin{equation}
F^{-1}(t,t_0) = \bar{T}\exp\left(i\int^t_{t_0} H_I(t)dt\right), \label{eq2-1-17}
\end{equation}
where $T$ ($\bar{T}$) represents (anti-)time ordering. 

A quantum operator ${\cal O}[\phi(t),\pi(t)]$ constructed from $\phi$
and $\pi$ can be written as
\begin{eqnarray}
{\cal O}(t) &=& F^{-1}(t,t_0){\cal O}^I(t) F(t,t_0) \nonumber\\
&=&\left[\bar{T}\exp\left(i\int^t_{t_0} H_I(t)dt\right)\right]{\cal O}^I(t)\left[T\exp\left(-i\int^t_{t_0} H_I(t)dt\right)\right], \label{eq2-1-18}
\end{eqnarray}
where ${\cal O}^I(t)\equiv {\cal O}[\phi^I(t),\pi^I(t)]$. Using
Eq.~\eqref{eq2-1-18}, we can compute the expectation value of the
operator $\langle{\cal O}(t)\rangle = \langle\Psi|{\cal
O}(t)|\Psi\rangle$ at a given time $t$ for an in-state
$|\Psi\rangle$ specified at the time $t_0$.  It is convenient to note
that
\begin{equation}
\langle{\cal O}(t)\rangle = \sum^{\infty}_{N=0}i^N\int^t_{t_0}dt_N\int^{t_N}_{t_0}dt_{N-1}\dots\int^{t_2}_{t_0}dt_1\langle[H_I(t_1),[H_I(t_2),\dots[H_I(t_N),{\cal O}^I(t)]\dots]]\rangle, \label{eq2-1-19}
\end{equation}
which can be derived by mathematical induction.

\subsection{\label{sec2-2}Mode expansion and occupation number}

Since the calculation in Eq.~\eqref{eq2-1-19} is performed in terms of
the interaction picture fields $\phi^I$ and $\pi^I$, it is convenient to
write down relevant quantities in the interaction picture.  From
Eqs.~\eqref{eq2-1-5} and \eqref{eq2-1-10}, the interaction picture
fields also satisfy the commutation relations,
\begin{align}
[\phi^I({\bf x},t),\pi^I({\bf y},t)] &= i\delta^{(3)}({\bf x-y}), \nonumber \\
[\phi^I({\bf x},t),\phi^I({\bf y},t)] &= [\pi^I({\bf x},t),\pi^I({\bf y},t)] = 0. \label{eq2-2-1}
\end{align}
Equations~(\ref{eq2-1-9}) imply that $\phi^I$ and $\pi^I$ are solutions of
free field equations of motion, giving their mode expansions,
\begin{align}
\phi^I({\bf x},t) &= \int \frac{d^3p}{(2\pi)^3}\frac{1}{\sqrt{2E_p}}\left[e^{ip\cdot x}a_{\bf p}^I + e^{-ip\cdot x}a_{\bf p}^{I\dagger} \right], \label{eq2-2-2}\\
\pi^I({\bf x},t) &= -i\int \frac{d^3p}{(2\pi)^3}\sqrt{\frac{E_p}{2}}\left[e^{ip\cdot x}a_{\bf p}^I - e^{-ip\cdot x}a_{\bf p}^{I\dagger} \right], \label{eq2-2-3}
\end{align}
where $E_p=\sqrt{m^2+|{\bf p}|^2}$, $x^0=t$, and $p^0=E_p$.  Then, the
commutation relations~\eqref{eq2-2-1} are equivalent to
\begin{equation}
[a_{\bf p}^I,a_{\bf p'}^{I\dagger}] = (2\pi)^3\delta^{(3)}({\bf p}-{\bf p}'), \quad\mathrm{and}\quad [a_{\bf p}^I,a_{\bf p'}^I] = [a_{\bf p}^{I\dagger},a_{\bf p'}^{I\dagger}] = 0. \label{eq2-2-4}
\end{equation}

The creation and annihilation operators diagonalize the free Hamiltonian
of interaction picture fields,
\begin{equation}
H_0[\phi^I(t),\pi^I(t)] =\int d^3 x\left[\frac{1}{2}(\pi^I)^2+\frac{1}{2}(\nabla\phi^I)^2+\frac{1}{2}m^2(\phi^I)^2\right]
= \int\frac{d^3 p}{(2\pi)^3}E_p\left(a_{\bf p}^{I\dagger}a_{\bf p}^I+\frac{1}{2}(2\pi)^3\delta^{(3)}(0)\right). \label{eq2-2-5}
\end{equation}
Here, let us define the operator whose eigenvalue gives the occupation
number of a momentum state ${\bf p}$,
\begin{equation}
\hat{{\cal N}}_{\bf p} \equiv \frac{d^3 p}{(2\pi)^3}a_{\bf p}^{I\dagger}a_{\bf p}^I. \label{eq2-2-6}
\end{equation}
On the other hand, its eigenstate can be obtained by applying the ladder
operator $a_{\bf p}^{I\dagger}$ on the vacuum state defined by
\begin{equation}
a^I_{\bf p}|0\rangle_I = 0. \label{eq2-2-7}
\end{equation}

An operator similar to~\eqref{eq2-2-6} in the Heisenberg picture can
also be constructed. Since the Heisenberg picture and interaction
picture operators are related,
\begin{equation}
\phi({\bf x},t) = F^{-1}(t,t_0)\phi^I({\bf x},t)F(t,t_0) \quad\mathrm{and}\quad \pi({\bf x},t) = F^{-1}(t,t_0)\pi^I({\bf x},t)F(t,t_0), \label{eq2-2-8}
\end{equation}
the following time-dependent operators are useful:
\begin{equation}
a_{\bf p}(t) = F^{-1}(t,t_0)a_{\bf p}^IF(t,t_0) \quad \mathrm{and} \quad a_{\bf p}^{\dagger}(t) = F^{-1}(t,t_0)a_{\bf p}^{I\dagger}F(t,t_0). \label{eq2-2-9}
\end{equation}
From Eqs.~(\ref{eq2-2-4}) and (\ref{eq2-2-9}), it is manifest that
$a_{\bf p}(t)$ and $a_{\bf p}^{\dagger}(t)$ also satisfy the canonical
commutation relations, and diagonalize the free Hamiltonian of
Heisenberg picture fields $H_0[\phi(t),\pi(t)]$.  Hence we recognize
that the operator
\begin{equation}
\hat{{\cal N}}_{\bf p}(t) \equiv \frac{d^3 p}{(2\pi)^3}a_{\bf p}^{\dagger}(t)a_{\bf p}(t) \label{eq2-2-10}
\end{equation}
describes the time evolution of the occupation number, and we substitute
it into ${\cal O}(t)$ in the left-hand side of Eq.~\eqref{eq2-1-19}.
However, in the actual calculation, the operator~\eqref{eq2-2-6} is used
as ${\cal O}^I$ in the right-hand side of Eq.~\eqref{eq2-1-19}.

Note that neither $a^I_{\bf p}$ nor $a_{\bf p}(t)$ diagonalizes the full
Hamiltonian $H=H_0+H_I$, and the state $|0\rangle_I$ given by
Eq.~\eqref{eq2-2-7} is not the ground state of the full Hamiltonian,
which we shall denote $|0\rangle_H$.  On the other hand, the in-state,
which is used to calculate the expectation value in
Eq.~\eqref{eq2-1-19}, is defined as an eigenstate of $H$, not
$H_0$. Such a state can be constructed by applying the operator
$a^{\mathrm{in}\dagger}_{\bf p}$, which creates one particle state from
$|0\rangle_H$.  It is assumed that this in-state approaches a free
particle state constructed by $a^{\dagger}(t)$ in the limit $t_0-t\to
-\infty$, up to a factor representing the renormalization of the wave
function. Such a factor can be absorbed into the physical mass of the
field $\phi$, which differs from the bare mass $m$ appealing in
$H_0$~\cite{Peskin:1995ev}.

Aside from the renormalization factor, in the limit $t_0-t\to -\infty$,
the right hand side of Eq.~\eqref{eq2-1-19} can be reduced into the
expectation value in the ``vacuum'' $|0\rangle_I$ multiplied by a factor
arising from the overlap between states $|0\rangle_I$ and $|0\rangle_H$.
This overlapping factor drops out when we divide the expectation value
by $1= _H\langle 0|0\rangle_H$.  This procedure is justified by taking
the limit $t_0-t\to-\infty(1-i\epsilon)$ in a slightly imaginary
direction, where $\epsilon$ is a positive infinitesimal. Note that, in
this case, the quantity appearing in the denominator is equal to unity
because of $_I\langle0|F^{-1}(t,t_0)F(t,t_0)|0\rangle_I = 1$.  This fact
implies that all vacuum fluctuations automatically vanish in the in-in
formalism~\cite{Weinberg:2005vy}.

Having removed ambiguities via the procedure described above, we can
simply calculate the right-hand side of Eq.~\eqref{eq2-1-19} with the
state constructed by applying operators $a^{I\dagger}_{\bf p}$ on the
vacuum state $|0\rangle_I$ in the interaction picture. This initial
state will be specified in the next subsection.

In the following, the subscript ``$I$'' is omitted for simplicity. It is
convenient to consider a finite spatial box with volume $V=L^3$ so that
the label of each mode becomes discrete, ${\bf p_n}=(2\pi/L){\bf n}$,
and ${\bf n}=(n_x,n_y,n_z)$, where $n_x$, $n_y$ and $n_z$ are integers.
Then we just take the following replacements:
\begin{align}
(2\pi)^3\delta^{(3)}({\bf p}-{\bf p}') &\to V\delta_{n_x,n_x'}\delta_{n_y,n_y'}\delta_{n_z,n_z'}, \nonumber \\
\int\frac{d^3p}{(2\pi)^3} &\to \frac{1}{V}\sum_{n_x,n_y,n_z}, \nonumber
\end{align}
and also
\begin{equation}
[a_i,a_j^{\dagger}] = V\delta_{i,j},\quad\mathrm{and}\quad [a_i,a_j]=[a_i^{\dagger},a_j^{\dagger}] = 0. \label{eq2-2-11}
\end{equation}
Here, the italic indices $i$, $j$ should be understood as abbreviated
notation for three dimensional vectors with discrete components.  The
number operator defined in Eq.~\eqref{eq2-2-6} is rewritten as
\begin{equation}
\hat{{\cal N}}_n \equiv \frac{a_n^{\dagger}a_n}{V}, \label{eq2-2-12}
\end{equation}
where the factor $1/V$ appears due to the factor $V$ in
Eq.~\eqref{eq2-2-11}.

\subsection{\label{sec2-3}Coherent oscillation as a quantum state}

Now let us specify the state at the initial time. Since we are
interested in the evolution of coherently oscillating axions, the
initial time is set to be the epoch of the QCD phase transition, at
which the mass $m$ becomes greater than the Hubble parameter $H$ and the
classical axion field begins to oscillate around the minimum of the
potential. These axions are produced due to the misalignment
mechanism~\cite{Preskill:1982cy}, and are called the zero mode, since a
huge number of axions homogeneously oscillate over a large distance.

In addition to this zero mode, however, there are additional
contributions to the axion abundance. One contribution is produced by
the thermal bath in the early Universe, and its abundance is fixed at
the decoupling temperature~\cite{Masso:2002np}.  Another contribution
comes from the decay of topological defects, such as global strings and
domain walls~\cite{Davis:1986xc,Lyth:1991bb}.  Both of them have
definite momenta, and we call them the nonzero modes in contrast to the
zero mode.  If inflation occurred before the Peccei-Quinn (PQ) phase
transition, those produced by topological defects can be a dominant
component of dark matter.  On the other hand, if inflation occurred
after the PQ phase transition, their population is negligible.
See~\cite{Hiramatsu:2010yu,Hiramatsu:2012gg,Hiramatsu:2012sc} for recent
developments about this issue.

Each of the zero mode and the nonzero modes correspond to a definite
quantum state.  In particular, it is possible to construct a state with
a definite momentum ${\bf p_k}$ occupied by ${\cal N}_k$ axions as a
{\it number state},
\begin{equation}
|{\cal N}_k\rangle = \frac{1}{\sqrt{{\cal N}_k!V^{{\cal N}_k}}}(a_k^{\dagger})^{{\cal N}_k}|0\rangle_I, \label{eq2-3-1}
\end{equation}
where $|0\rangle_I$ is the vacuum defined by Eq.~\eqref{eq2-2-7}.  This
is an eigenstate of the number operator~\eqref{eq2-2-12}, and we can
construct complete orthonormal basis by using a series of the number
states. Here, it should be noted that the nonzero modes correspond to
the number states. In the classical limit, these states can be
interpreted as classical point particles with definite energies and
momenta.

On the other hand, the zero mode has different properties compared with
nonzero modes. It has a huge occupation number as large as ${\cal
N}\sim 10^{61}$. In the classical limit, this state should be
interpreted as a classical field, rather than point
particles~\cite{Erken:2011dz}. Such a highly degenerate Bose gas of
axions might be described as a {\it coherent
state}~\cite{1990A&A...231..301B}. Mathematically, a coherent state can
be represented by using the basis of number states~\cite{Glauber:1963tx},
\begin{equation}
|\alpha_i\rangle = e^{-\frac{1}{2}|\alpha_i|^2}\sum^{\infty}_{n=0}\frac{\alpha_i^n}{n!\sqrt{V^n}}(a^{\dagger}_i)^n|0\rangle_I, \label{eq2-3-2}
\end{equation}
where $\alpha_i$ is a complex number and the numerical factor is chosen
so that it is normalized $\langle\alpha_i|\alpha_i\rangle=1$. The
coherent state is characterized as an eigenvector of the annihilation
operator such that
\begin{equation}
a_i|\alpha_j\rangle = V^{1/2}\alpha_j\delta_{ij}|\alpha_j\rangle. \label{eq2-3-3}
\end{equation}

The coherent state representation is suitable for the modes outside the horizon at the time of the QCD phase transition.
This is because such modes start to oscillate simultaneously at that time and are not dephased.
We note that it is not an exact statement since the onset of the oscillation may vary for each of patches separated by the QCD horizon.
Such a difference is at most the order of the magnitude of temperature fluctuations, because the onset of the oscillation is determined by the 
temperature dependent axion mass [see Eq.~\eqref{eq3-1}].
Therefore, the difference in the onset of the oscillation can be simply ignored as long as we consider the transitions in the total number of axions.
For this reason, we expect that the modes outside the QCD horizon are well approximated as coherent states.

However, the value of the initial amplitude of the oscillation (called the initial misalignment angle) might be different for each of the QCD patches,
if the PQ phase transition occurs after inflation.
Furthermore, even for the case where the PQ phase transition occurs before inflation, the value of the initial misalignment angle might vary for some length scales
beyond the size of inflated patches.
Therefore, we expect that the coherently oscillating axions have momenta comparable to or less than the Hubble scale at the time of the QCD phase transition.
In this sense, the ``zero mode" is not exactly a single mode with zero momentum, but the collection of plural modes near the ground state.

Axions produced from the dynamics inside the horizon such as the decay of topological defects and the interaction with the thermal bath
are almost dephased, and they are not described as a coherent state.
Here, we simply assume that they are described as a number state given by Eq.~\eqref{eq2-3-1}.
In opposition to the zero modes, these modes have momenta larger than the Hubble scale at the time of the QCD phase transition.
We summarize the contents of the initial state in Table~\ref{tab1}.

Hereafter we assume that a huge number of particles occupy a small
number $K$ of states whose momenta are less than the Hubble scale at the time of QCD phase transition,
and that they are described as coherent states.
There also exist nonzero modes, which occupy states
with higher momenta and are described as number states. The collection
of such states can be expressed as
\begin{align}
|\{{\cal N}\},\{\alpha\}\rangle &= \prod_{k>K}\frac{1}{\sqrt{{\cal N}_k!V^{{\cal N}_k}}}(a_k^{\dagger})^{{\cal N}_k}|\{\alpha\}\rangle, \label{eq2-3-4}\\
|\{\alpha\}\rangle &= \prod_{i\le K}e^{-\frac{1}{2}|\alpha_i|^2}\sum_{n=0}^{\infty}\frac{\alpha_i^n}{n!\sqrt{V^n}}(a_i^{\dagger})^n|0\rangle_I. \label{eq2-3-5}
\end{align}
Note that $i\le K$ is the abbreviated notation representing the sum over
lowest $K$ modes (i.e.~actual states are labeled by three momenta, and
we must distinguish them by spatial directions of momenta as well as
their absolute value).  Let us call the modes with $i>K$ the
particlelike modes and the modes with $i\le K$ the condensed modes.
It would be convenient to note the following relations:
\begin{align}
\left[a_i,(a_j^{\dagger})^{{\cal N}_j}\right] &= {\cal N}_jV\delta_{ij}(a^{\dagger}_j)^{{\cal N}_j-1}, \label{eq2-3-6}\\
\left[(a_i)^{{\cal N}_i},a_i^{\dagger}\right] &= {\cal N}_jV\delta_{ij}(a_j)^{{\cal N}_j-1}, \label{eq2-3-7}\\
a_k|\{{\cal N}\},\{\alpha\}\rangle &= 
\left\{
\begin{array}{lll}
\sqrt{{\cal N}_kV}|\{{\cal N}\}^k,\{\alpha\}\rangle & \mathrm{if} & k>K\\
\alpha_k\sqrt{V}|\{{\cal N}\},\{\alpha\}\rangle & \mathrm{if} & k\le K\\
\end{array}
\right.,
\label{eq2-3-8}
\end{align}
where $|\{{\cal N}\}^k,\{\alpha\}\rangle$ is the state obtained by
replacing the factor $(a_k^{\dagger})^{{\cal N}_k}/\sqrt{{\cal
N}_k!V^{{\cal N}_k}}$ with $(a_k^{\dagger})^{{\cal N}_k-1}/\sqrt{({\cal
N}_k-1)!V^{{\cal N}_k-1}}$ in Eq.~\eqref{eq2-3-4}.

It is important to assume that there are plural condensed modes ($K>1$).
Our interest is to know how these condensed modes reach thermal
equilibrium by exchanging their momenta.
If condensed modes never thermalize, their occupation number does not change from that of the initial states where
a number of particles occupy plural states labeled by momenta comparable to or less than the Hubble scale at the time of QCD phase transition.
However, once the effects of self interaction become relevant, transition between condensed modes rapidly occurs.
Then, the initial distribution begins to change toward the equilibrium form. 

\begin{table}[h]
\begin{center} 
\caption{Classification of axions with their origins and quantum state representations.}
\vspace{3mm}
\begin{tabular}{c c c}
\hline\hline
 & Production mechanism & Quantum state \\
\hline
Zero mode & Misalignment mechanism & Coherent states (condensed modes) \\
Nonzero mode (topological defects) & Decay of defects & Number states (particlelike modes) \\
Nonzero mode (thermal axions) & Thermal decoupling & Number states (particlelike modes) \\
\hline\hline
\label{tab1}
\end{tabular}
\end{center}
\end{table}

Let us take the expectation value of $\phi$ given by Eq.~\eqref{eq2-2-2}
at the initial time $t_0$ for the state $|\{{\cal
N}\},\{\alpha\}\rangle$,
\begin{align}
\phi_0 &\equiv \langle\{{\cal N}\},\{\alpha\}|\phi({\bf x}, t_0)|\{{\cal N}\},\{\alpha\}\rangle \nonumber\\
&= \sum_{n\le K}\frac{1}{\sqrt{2E_n V}}(e^{-iE_nt_0+i{\bf p_n\cdot x}}\alpha_n + e^{iE_nt_0-i{\bf p_n\cdot x}}\alpha_n^*). \label{eq2-3-9}
\end{align}
Since the wavelength of the condensed modes is comparable or greater
than the QCD horizon, $|{\bf p_n}|\lesssim H(t_0)\sim t^{-1}_0$, ${\bf
p_n\cdot x} \ll 1$ and hence the factor $e^{\pm i{\bf p_n\cdot x}}$ is
negligible. This approximation remains valid as long as we consider the
dynamics inside the horizon.  We also approximate
$E_n=\sqrt{m^2+p_n^2}\simeq m$ since the coherent oscillation begins
when $|{\bf p_n}|\lesssim H(t_0) < m$ is satisfied. Then, the
expectation value, $\phi_0$, is given by
\begin{align}
\phi_0 &\simeq \sum_{n\le K}\frac{1}{\sqrt{2mV}}(e^{-imt_0}\alpha_n+e^{imt_0}\alpha_n^*) \nonumber\\
&= \sum_{n\le K} \sqrt{\frac{2}{mV}}|\alpha_n|\cos(mt_0-\beta_n), \label{eq2-3-10}
\end{align}
with
\begin{equation}
\alpha_n=|\alpha_n|e^{i\beta_n}. \label{eq2-3-11}
\end{equation}
If the condensed modes are decoupled with each other, the expectation
value of the field oscillates like $\langle\phi\rangle
\propto\cos(mt-\beta_n)$. Each mode oscillates independently with
different amplitude $|\alpha_n|$ and the total amplitude is given by the
superposition of $K$ oscillating modes.

Next, let us take the mean square deviation of the field amplitude for a
single coherent state given in Eq.~\eqref{eq2-3-2},
\begin{align}
\Delta\phi &= \sqrt{\langle\alpha_i|\phi^2|\alpha_i\rangle - \langle\alpha_i|\phi|\alpha_i\rangle^2} \nonumber\\
&= \sqrt{\frac{1}{V}\sum_n\frac{1}{2E_n}} \xrightarrow{V\to\infty}\sqrt{\int\frac{d^3p}{(2\pi)^3}\frac{1}{2E_p}}. \label{eq2-3-12}
\end{align}
Since this result does not depend on $\alpha_i$, it holds for the state
with $\alpha_i=0$ (the vacuum state), which implies that the deviation
given in Eq.~\eqref{eq2-3-12} is nothing but the vacuum
fluctuation. Therefore, the coherent state has the same trajectory with
the classical field and the same fluctuation with the vacuum.

The expectation value of the momentum conjugate~\eqref{eq2-2-3} leads to
\begin{align}
\dot{\phi}_0 &\equiv \langle\{{\cal N}\},\{\alpha\}|\pi({\bf x}, t_0)|\{{\cal N}\},\{\alpha\}\rangle \nonumber\\
&= \sum_{n\le K}\sqrt{\frac{2m}{V}}|\alpha_n|\sin(\beta_n-mt_0). \label{eq2-3-13}
\end{align}
Let us assume that the initial velocity $\langle\dot{\phi}\rangle$ of every
mode vanishes, $\beta_n=mt_0$. In this case, we obtain
\begin{equation}
\phi_0 \simeq \sum_{n\le K}\sqrt{\frac{2}{mV}}|\alpha_n| = \sum_{n\le K}\theta^{\mathrm{ini}}_{n}F_a, \label{eq2-3-14}
\end{equation}
where $F_a$ is the axion decay constant and
\begin{equation}
\theta^{\mathrm{ini}}_n \equiv \sqrt{\frac{2}{mV}}\frac{|\alpha_n|}{F_a} \label{eq2-3-15}
\end{equation}
is the initial misalignment angle for a mode $n$.

The total number of particles at the initial time is given by
\begin{equation}
N = \sum_n\langle\{{\cal N}\},\{\alpha\}|\hat{\cal N}_n|\{{\cal N}\},\{\alpha\}\rangle, \label{eq2-3-16}
\end{equation}
where $\hat{{\cal N}}_n$ is the number operator given in
Eq.~\eqref{eq2-2-12}.  Dividing it by a volume $V$ yields the number
density of axions
\begin{equation}
n_{\mathrm{tot}} = \frac{N}{V} = \frac{1}{V^2}\sum_n \langle\{{\cal N}\},\{\alpha\}|a_n^{\dagger}a_n|\{{\cal N}\},\{\alpha\}\rangle = n_p + n_c, \label{eq2-3-17}
\end{equation}
where
\begin{equation}
n_p \equiv \frac{1}{V}\sum_{n>K}{\cal N}_n \label{eq2-3-18}
\end{equation}
is the number density of particlelike modes, and
\begin{equation}
n_c\equiv \sum_{n\le K}n_{c,n} = \frac{1}{2}mF_a^2(\theta^{\mathrm{ini}})^2,\quad n_{c,n}\equiv \frac{1}{V}|\alpha_n|^2 \label{eq2-3-19}
\end{equation}
are the number densities of condensed modes. Here
$(\theta^{\mathrm{ini}})^2$ is the square of the total misalignment
angle
\begin{equation}
(\theta^{\mathrm{ini}})^2 \equiv \sum_{n\le K}(\theta^{\mathrm{ini}}_n)^2. \label{eq2-3-20}
\end{equation}
In the continuous limit $V\to \infty$, Eq.~\eqref{eq2-3-17} can be
rewritten as
\begin{equation}
n_{\mathrm{tot}} = \int \frac{d^3p}{(2\pi)^3}f({\bf p}), \label{eq2-3-21}
\end{equation}
where $f({\bf p})$ is the total phase space distribution function of
axions,
\begin{equation}
f({\bf p}) = {\cal N}_{\bf p} + \sum_{n\le K}(2\pi)^3\delta^{(3)}({\bf p-p_n})n_{c,n}. \label{eq2-3-22}
\end{equation}

\subsection{\label{sec2-4}Time evolution of quantum occupation number}

In the following, we compute the time evolution of the expectation value
of the quantum number operator~\eqref{eq2-1-19}.  By using the in-in
formalism, the time evolution of the occupation number is given by
\begin{equation}
\langle{\cal \hat{N}}_p(t)\rangle = \langle {\cal \hat{N}}_p \rangle + i\int^t_{t_0}dt_1\langle[H_I(t_1),{\cal \hat{N}}_p]\rangle
 - \int^t_{t_0}dt_2\int^{t_2}_{t_0}dt_1\langle[H_I(t_1),[H_I(t_2),{\cal \hat{N}}_p]]\rangle + (\mathrm{higher\ order\ in}\ H_I), \label{eq2-4-1}
\end{equation}
where $\langle\dots\rangle$ represents the expectation value for the
state given by Eq.~\eqref{eq2-3-4}. We consider the following form of
the interaction~\cite{Erken:2011dz},
\begin{equation}
H_I(t) = \frac{1}{V^4}\sum_{ijkl}\frac{1}{4}\Lambda^{ij}_{kl}e^{-i\Omega^{ij}_{kl}t}a_k^{\dagger}a_l^{\dagger}a_ia_j, \label{eq2-4-2}
\end{equation}
where $\Omega^{ij}_{kl}\equiv E_i+E_j - E_k- E_l$ and
$\Lambda^{ij}_{kl}$ satisfies
$\Lambda^{ij}_{kl}=\Lambda^{ji}_{kl}=\Lambda^{ij}_{lk}=\Lambda^{kl*}_{ij}$.
This can be obtained from $\lambda\phi^4/4!$ type interaction in the
effective Lagrangian of the axion field with $\lambda\simeq
0.35m^2/F_a^2$, and the coefficient $\Lambda^{ij}_{kl}$ becomes
\begin{equation}
\Lambda^{\ ij}_{s\ \ kl} = -\frac{\lambda}{4\sqrt{E_iE_jE_kE_l}}V\delta_{i+j,k+l}. \label{eq2-4-3}
\end{equation}
Here we dropped the processes which violate axion number such as
$a^{\dagger}a^{\dagger}a^{\dagger}a$, since
we are interested in the term of first order in $H_I$, where
axion number violating processes are forbidden due to the conservation of energy and three momenta
if axions are nonrelativistic.
Such axion number violating self-interaction terms are relevant only in the higher order in perturbation theory.
The possibility of the axion number violating process including other particle species will be discussed in the next subsection.
In addition to the self coupling, axions
also interact due to their gravitational potential.  In the Newtonian
limit, the interaction Hamiltonian of the gravitational coupling is
given by
\begin{equation}
H_{I,g}[\phi(t),\pi(t)] = -\frac{G}{2}\int d^3xd^3x'\frac{\rho({\bf x},t)\rho({\bf x}',t)}{|{\bf x-x}'|}, \label{eq2-4-4}
\end{equation}
where $G$ is the Newton's constant and $\rho({\bf x},t)=(\pi^2({\bf
x},t)+m^2\phi^2({\bf x},t))/2$ is the energy density of axions.  This
leads to the term~\eqref{eq2-4-2} with the coefficient
\begin{equation}
\Lambda^{\ ij}_{g\ \ kl} = -4\pi G m^2\left(\frac{1}{|{\bf p}_k-{\bf p}_i|^2}+\frac{1}{|{\bf p}_k-{\bf p}_j|^2}\right)V\delta_{i+j,k+l}, \label{eq2-4-5}
\end{equation}
where we used the approximation $E_i\approx m$.

Let us evaluate the term of first order in $H_I$. Due to the following
relation,
\begin{equation}
[a_k^{\dagger}a^{\dagger}_la_ia_j,a^{\dagger}_pa_p] = V\delta_{ip}a^{\dagger}_ka^{\dagger}_la_ja_p + V\delta_{jp}a^{\dagger}_ka^{\dagger}_la_ia_p
- V\delta_{kp}a^{\dagger}_pa^{\dagger}_la_ia_j - V\delta_{lp}a^{\dagger}_pa^{\dagger}_ka_ia_j, \label{eq2-4-6}
\end{equation}
the commutation relation becomes
\begin{equation}
[H_I(t),\hat{{\cal N}}_p] = \frac{1}{2V^4}\sum_{jkl}\left[\Lambda^{pj}_{kl}e^{-i\Omega^{pj}_{kl}t}a^{\dagger}_ka^{\dagger}_la_ja_p - \mathrm{H.c.}\right]. \label{eq2-4-7}
\end{equation}
The following relations coming from Eq.~\eqref{eq2-3-8} are useful to
take the expectation value of the above term,
\begin{equation}
a_{k'}a_k |\{{\cal N}\},\{\alpha\}\rangle =
\left\{
\renewcommand{\arraystretch}{1.8}
\begin{array}{l l l}
\sqrt{{\cal N}_k({\cal N}_k-1)}V|\{{\cal N}\}^{2k}, \{\alpha\} \rangle & \mathrm{if} & k=k'>K\\
\sqrt{{\cal N}_k{\cal N}_{k'}}V|\{{\cal N}\}^{k,k'}, \{\alpha\} \rangle & \mathrm{if} & k\ne k',\ k>K,\ \mathrm{and}\ k'> K \\
\sqrt{{\cal N}_k}\alpha_{k'}V|\{{\cal N}\}^k, \{\alpha\} \rangle & \mathrm{if} & k>K\ \mathrm{and}\ k'\le K \\
\alpha_k\alpha_{k'}V|\{{\cal N}\}, \{\alpha\} \rangle & \mathrm{if} & k\le K,\ \mathrm{and}\ k'\le K 
\end{array}
\renewcommand{\arraystretch}{1}
\right., \label{eq2-4-8}
\end{equation}
where the state $|\{{\cal N}\}^{2k}, \{\alpha\} \rangle$ contains a
factor $(a^{\dagger}_k)^{{\cal N}_k-2}/\sqrt{({\cal N}_k-2)!V^{{\cal
N}_k-2}}$ for mode $k$, and the state $|\{{\cal N}\}^{k,k'}, \{\alpha\}
\rangle$ contains a factor $(a^{\dagger}_k)^{{\cal
N}_k-1}(a^{\dagger}_{k'})^{{\cal N}_{k'}-1}/\sqrt{({\cal
N}_k-1)!V^{{\cal N}_k-1}({\cal N}_{k'}-1)!V^{{\cal N}_{k'}-1}}$ for
modes $k$ and $k'$.  By separating the summation over indices $jkl$ into
the contribution of particlelike modes $>K$ and that of condensed modes
$\le K$, and using Eq.~\eqref{eq2-4-8}, we can compute the expectation
value of Eq.~\eqref{eq2-4-7}. For $p\le K$, after some algebra, we
obtain
\begin{align}
\langle[H_I(t),\hat{{\cal N}}_p]\rangle = &\frac{1}{V^2}\sum_{j>K}\sum_{k\le K}\left[\Lambda^{pj}_{kj}e^{-i(E_p-E_k)t}{\cal N}_j\alpha_p\alpha_k^* - \mathrm{c.c.}\right] \nonumber \\
& + \frac{1}{2V^2}\sum_{j\le K}\sum_{k\le K}\sum_{l\le K}\left[\Lambda^{pj}_{kl}e^{-i\Omega^{pj}_{kl}t}\alpha_k^*\alpha_l^*\alpha_j\alpha_p - \mathrm{c.c.}\right]
\quad\mathrm{for}\quad p\le K. \label{eq2-4-9}
\end{align}
The first term in the right-hand side of Eq.~\eqref{eq2-4-9} vanishes
since $\Lambda^{pj}_{kj}$ contains the conservation law of three momenta
$\delta_{p+j,k+j}$.  On the other hand, for $p>K$, this term
exactly vanishes
\begin{equation}
\langle[H_I(t),\hat{{\cal N}}_p]\rangle = 0 \quad \mathrm{for} \quad p>K. \label{eq2-4-10}
\end{equation}
Finally, taking the time integration yields the contribution at the
first order perturbation,
\begin{align}
i\int^t_{t_0}dt_1\langle[H_I(t),\hat{{\cal N}}_p]\rangle &= 
\frac{1}{2V^2}\sum_{j\le K}\sum_{k\le K}\sum_{l\le K}\left[\Lambda^{pj}_{kl}\frac{e^{-i\Omega^{pj}_{kl}t}(e^{i\Omega^{pj}_{kl}(t-t_0)}-1)}{\Omega^{pj}_{kl}}\alpha_k^*\alpha_l^*\alpha_j\alpha_p +\mathrm{c.c.}\right] \nonumber\\
&\simeq -\frac{1}{2V^2}\sum_{j\le K}\sum_{k\le K}\sum_{l\le K}\left[\Lambda^{pj}_{kl}\frac{e^{-i\Omega^{pj}_{kl}t}}{\Omega^{pj}_{kl}}\alpha_k^*\alpha_l^*\alpha_j\alpha_p +\mathrm{c.c.}\right]
\quad\mathrm{for}\quad p\le K, \label{eq2-4-11}
\end{align}
and
\begin{equation}
i\int^t_{t_0}dt_1\langle[H_I(t),\hat{{\cal N}}_p]\rangle = 0 \quad \mathrm{for} \quad p>K, \label{eq2-4-12}
\end{equation}
where we have dropped the rapidly oscillating term
$e^{i\Omega^{pj}_{kl}(t-t_0)}$ as $t-t_0\to \infty$ in the second line
of Eq.~\eqref{eq2-4-11}.
Note that, if there is no scattering ($p=k$ or
$p=l$), the first line of Eq.~\eqref{eq2-4-11} also vanishes.

As was conjectured in~\cite{Erken:2011dz}, there are two distinct
regimes for the interaction process.  One is the particle kinetic regime
characterized by the condition $\Gamma\ll\delta\omega$, where $\Gamma$
is the evolution rate of the system and $\delta\omega$ is the typical
energy exchanged in the interaction. In this regime we expect that
$\Omega^{pj}_{kl}t\gg 1$ and the factor $\exp(-i\Omega^{pj}_{kl}t)$ in
Eq.~\eqref{eq2-4-11} cancels out when the time average is taken. Hence
the first order term~\eqref{eq2-4-11} becomes irrelevant, which requires
us to evaluate second order terms in the expansion~\eqref{eq2-4-1} to
follow the time evolution of the occupation number. The explicit
calculation for second order terms is given in the Appendix.

The opposite regime characterized by the condition $\Gamma\gg
\delta\omega$ is called the condensed regime.  Since
$\Omega^{pj}_{kl}t\ll 1$ is satisfied, we can safely set
\begin{equation}
e^{-i\Omega^{pj}_{kl}t} \simeq 1 \label{eq2-4-13}
\end{equation}
in Eq.~\eqref{eq2-4-11}, and hence the first order term becomes relevant
for the estimation of the evolution rate.  In this regime, considerably
small momenta $\delta\omega$ are exchanged between $K$ highly occupied
states.  It makes sense even though $\Omega^{pj}_{kl}t\ll 1$, since the
transition occurs as ${\cal N}\Omega^{pj}_{kl}t\gg 1$ for huge number of
particles, ${\cal N}$. The expression~\eqref{eq2-4-11} will be used in
estimating the thermalization rate of axions in Sec.~\ref{sec3}.

Note that the approximation $e^{i\Omega^{pj}_{kl}(t-t_0)}\approx 0$ used in the second line of Eq.~\eqref{eq2-4-11}
is justified regardless of the condition of the condensed regime.
Here, we consider two time scales which should be distinguished from each other.
$(t-t_0)$ corresponds to the time scale in which the field cannot be interpreted as free field and the effect of the potential force should be introduced in order to describe its time evolution.
Hence the approximation $e^{i\Omega^{pj}_{kl}(t-t_0)}\approx 0$ follows from the asymptotic condition in which the in-state defined at $t_0$ approaches a free particle state in the limit $(t-t_0) \to \infty$,
as is assumed in the usual formulation of the quantum field theory [see discussions in the paragraph below Eq.~\eqref{eq2-2-10}].
However, we are interested in the time scale in which the transition between different momentum states occurs.
This corresponds to ``$t$," or a time scale $dt$ in which the change in the occupation number $d{\cal N}$
[i.e. the difference between ${\cal N}(t)$ and ${\cal N}(t+dt)$] becomes comparable with the total number of particles ${\cal N}$ in a state $p$, as defined in Eq.~\eqref{eq3-8}.
In other words, particles begin to feel the potential force in the time scale $t-t_0$, but they make a transition in the time scale $t\sim dt$.
Hence the condensed regime is defined by $\Omega^{pj}_{kl}t \ll 1$ rather than $\Omega^{pj}_{kl}(t-t_0) \ll 1$.
\subsection{\label{sec2-5}Interaction with other species}
So far we have considered only the self interaction of the axion field.
The highly degenerate axions may also couple with other species, such as
baryons, relativistic axions, and photons due to the gravitational
interactions, though the coupling with relativistic axions has already
been included in the formalism described in the previous subsection.  In
Ref.~\cite{Erken:2011dz}, it is claimed that axions have thermal contact
with other species after they form a BEC.  However, as shown in this
subsection, there are no such effects at least at the first order in the
perturbation theory.

In general, the interaction Hamiltonian with other species $b$ can be
written as
\begin{equation}
H_{I,b}(t) = \frac{1}{V^4}\sum_{ijkl}\frac{1}{4}\Lambda^{\ ij}_{b\ \ kl}e^{-i\Omega^{ij}_{kl}t}a_k^{\dagger}b_l^{\dagger}a_ib_j, \label{eq2-5-1}
\end{equation}
where $\Lambda^{\ ij}_{b\ \ kl}$ is a constant which contains the
conservation law of three momenta, and $b^{\dagger}_l$ and $b_l$ are the
operators which create and annihilate a species $b$ with the momentum
${\bf p}_l$, respectively. Here we consider only the interactions
conserving axion number at the leading order.
Interaction rates of the axion number violating processes
will be discussed later.
It should be also noticed that $b$ particles are conserved as well
at the vertex given by Eq.~\eqref{eq2-5-1}.

Substituting Eq.~\eqref{eq2-5-1} into Eq.~\eqref{eq2-4-1} enables us to
calculate how the occupation number of axions evolves with time due to
the interactions with other particles. The $b$ particles are assumed to
be in number states with a distribution ${\cal N}_{b,k}(t_0)$ at the
initial time,
\begin{equation}
|\{{\cal N}\},\{\alpha\},\{{\cal N}_b\}\rangle = \prod_k\frac{1}{\sqrt{{\cal N}_{b,k}!V^{{\cal N}_{b,k}}}}(b_k^{\dagger})^{{\cal N}_{b,k}}|\{{\cal N}\},\{\alpha\}\rangle, \label{eq2-5-2}
\end{equation}
where $|\{{\cal N}\},\{\alpha\}\rangle$ is given in Eq.~\eqref{eq2-3-4}.
${\cal N}_{b,k}$ can take a large number if $b$ is a boson, but it takes
either $0$ or $1$ if $b$ is a fermion.  In the following, the
state~\eqref{eq2-5-2} is taken in computing the expectation value in
Eq.~\eqref{eq2-4-1}. Noting that
\begin{equation}
[H_{I,b}(t),\hat{{\cal N}}_p] = \frac{1}{4V^4}\sum_{jkl}\left[\Lambda^{\ pj}_{b\ \ kl}e^{-i\Omega^{pj}_{kl}t}a^{\dagger}_kb^{\dagger}_la_pb_j - \mathrm{H.c.}\right], \label{eq2-5-3}
\end{equation}
we find for the condensed modes,
\begin{equation}
\langle[H_{I,b}(t),\hat{{\cal N}}_p]\rangle = \frac{1}{4V^2}\sum_j\sum_{k\le K}\left[\Lambda^{\ pj}_{b\ \ kj}e^{-i(E_p-E_k)t}{\cal N}_{b,j}\alpha_p\alpha_k^* - \mathrm{c.c.}\right]
\quad \mathrm{for} \quad p\le K, \label{eq2-5-4}
\end{equation}
which exactly vanishes because of the conservation law of three momenta
$\delta_{p+j,k+j}$ in $\Lambda^{\ pj}_{b\ \ kj}$.  This term also
vanishes for the particlelike modes,
\begin{equation}
\langle[H_{I,b}(t),\hat{{\cal N}}_p]\rangle = 0 \quad \mathrm{for} \quad p > K. \label{eq2-5-5}
\end{equation}
Hence there are no contributions from interactions with other species.

\begin{figure}[htp]
\centering
$\begin{array}{cc}
\subfigure[]{
\includegraphics[width=0.35\textwidth,angle=0]{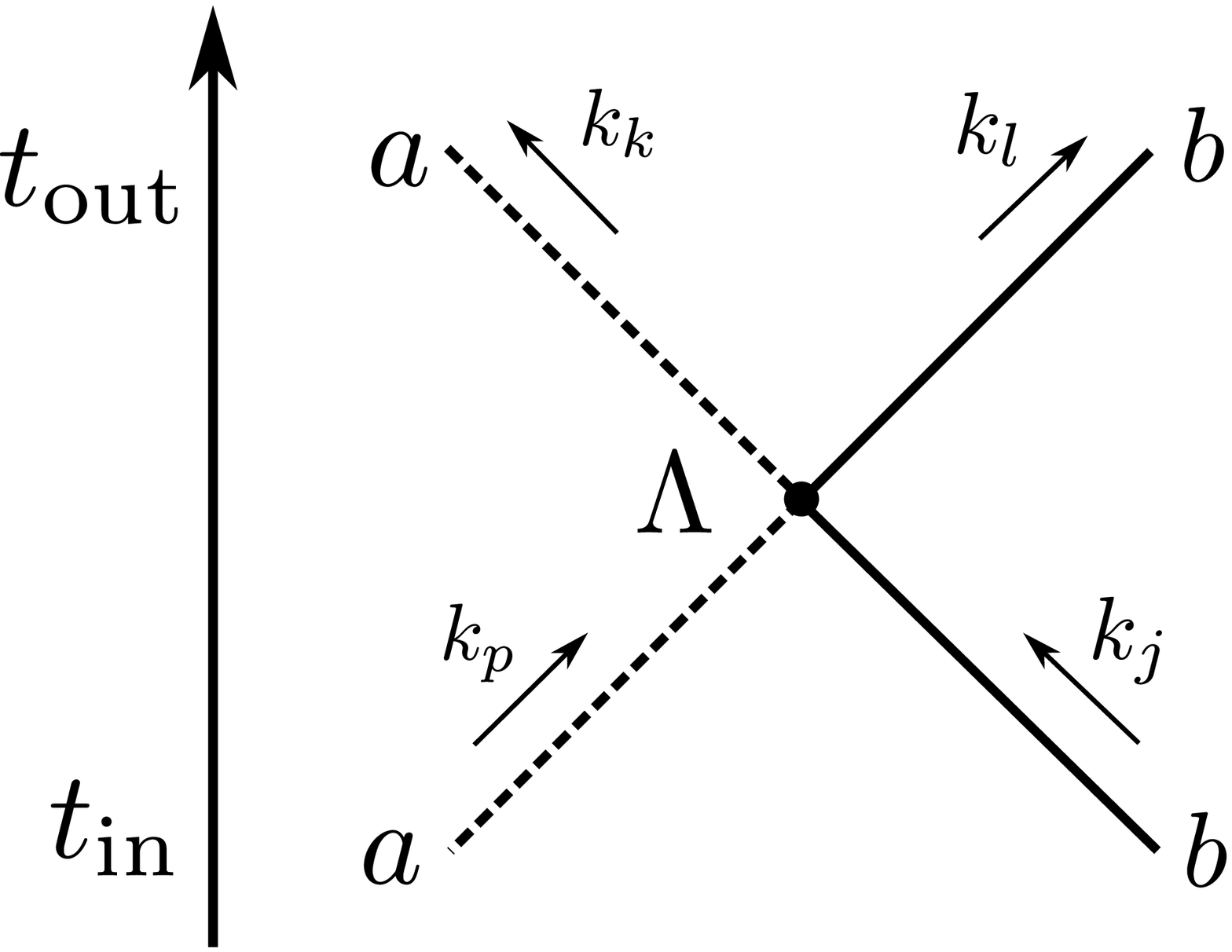}}
\hspace{40pt}
\subfigure[]{
\includegraphics[width=0.35\textwidth,angle=0]{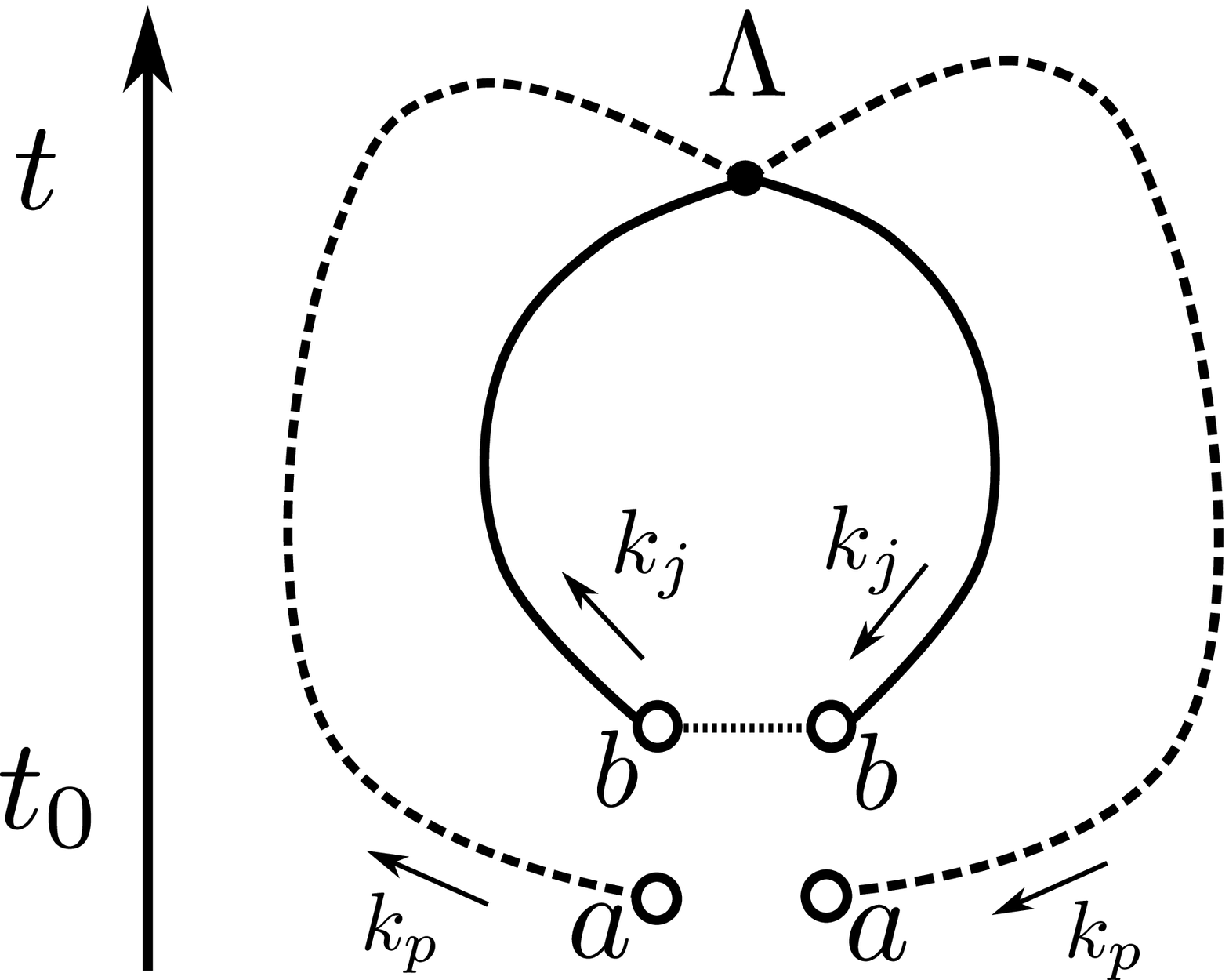}} \\
\subfigure[]{
\includegraphics[width=0.35\textwidth,angle=0]{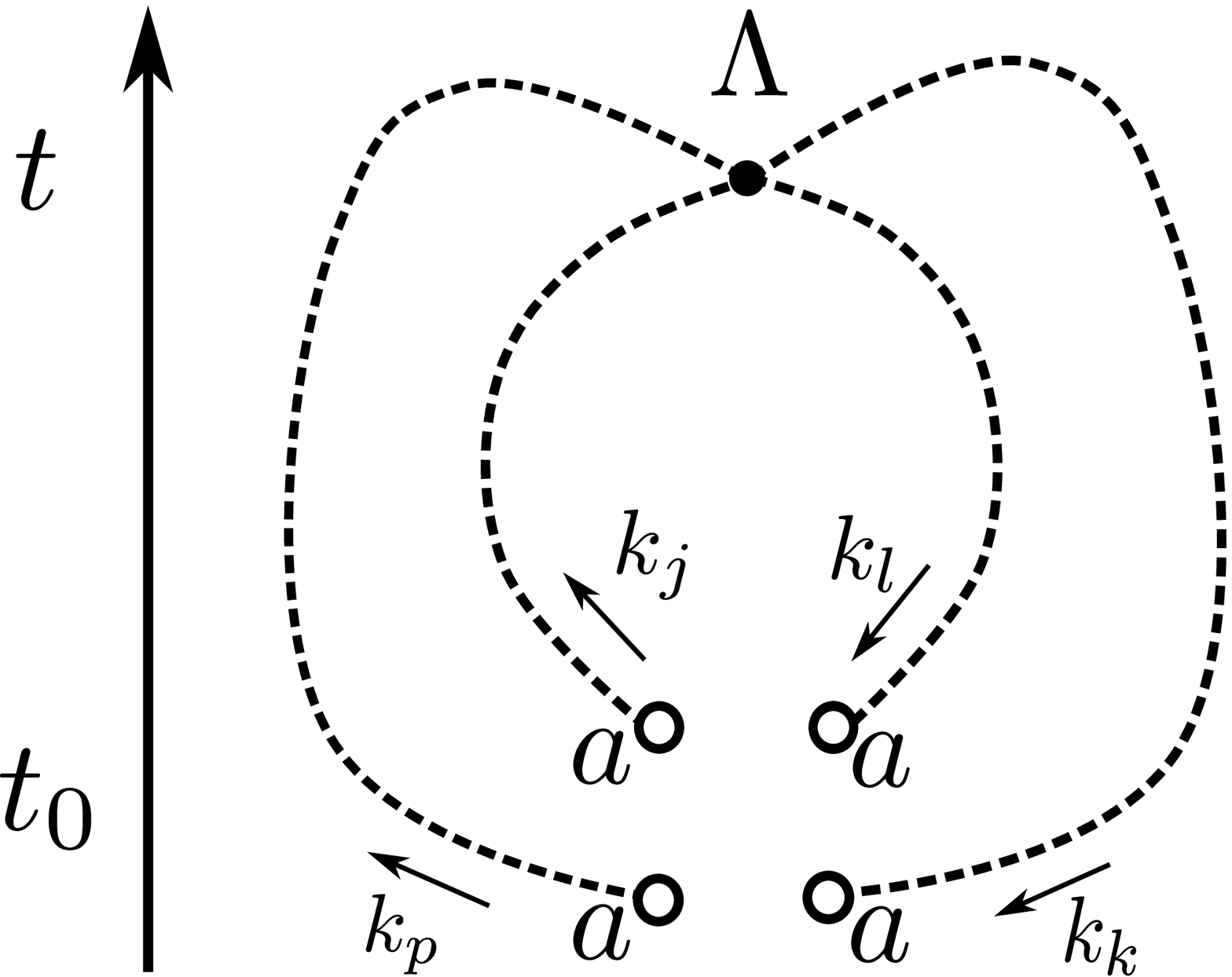}}
\hspace{40pt}
\subfigure[]{
\includegraphics[width=0.35\textwidth,angle=0]{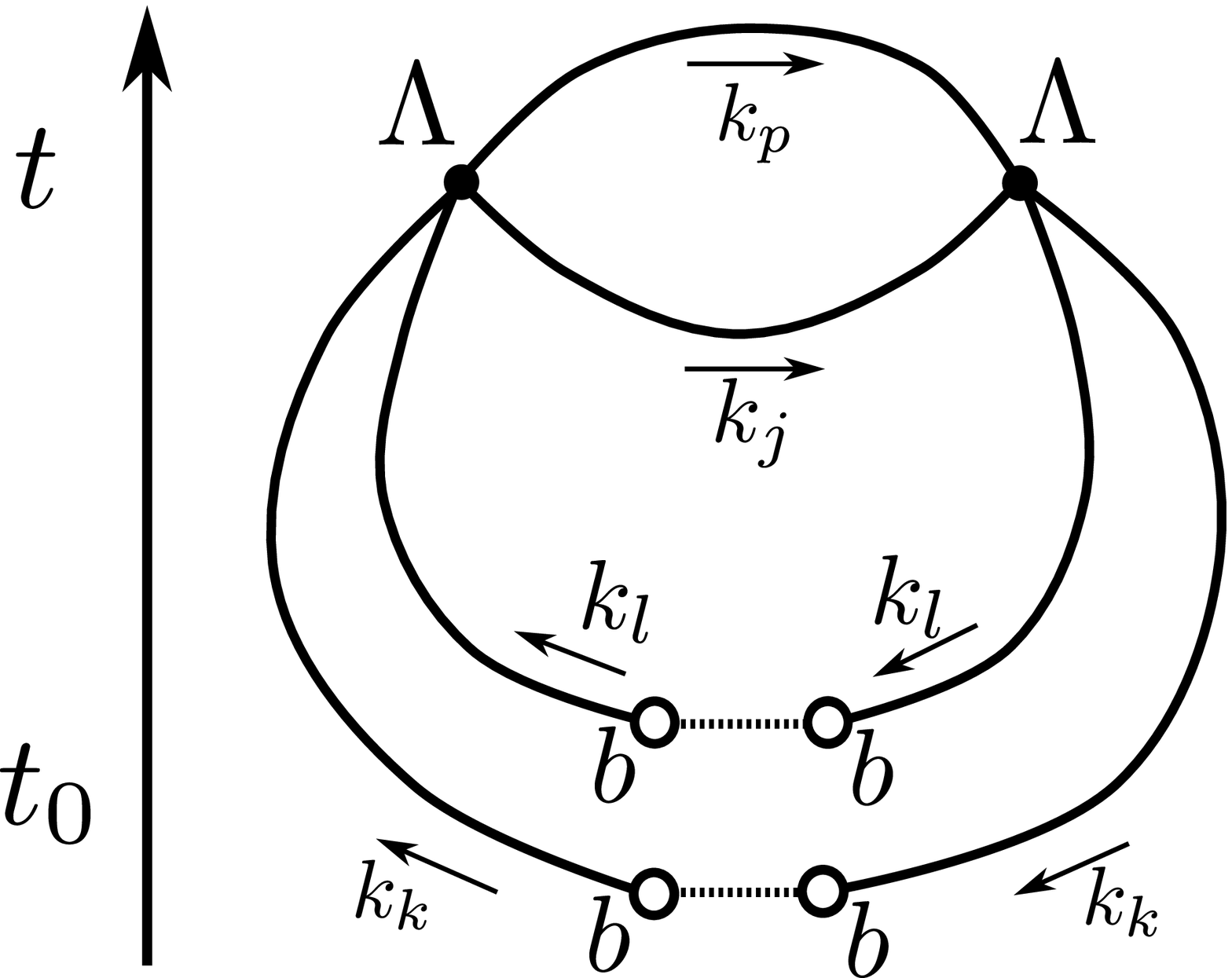}}
\end{array}$
\caption{Schematics of interaction processes. (a) The usual Feynmann
diagram for the tree level scattering process $a+b\to a+b$. The momentum
transfer occurs at the vertex denoted as $\Lambda$. (b) The tree level
diagram for the process $a+b\to a+b$ in the in-in formalism. If $b$ is a
particlelike mode, two momenta of in-states at $t=t_0$ must be
same. No momentum transfer occurs at the vertex.  (c) Tree level diagram
for the self interaction of condensed modes $a+a\to a+a$ in the in-in
formalism. Momenta of in-states can differ from each other.
Momentum transfer occurs at the vertex. (d) The diagram for the second
order scattering process between particlelike modes $b+b\to b+b$ in the
in-in formalism.  Momenta of in-states at $t=t_0$ are same, but
momentum transfer occurs at each vertex.}  \label{fig1}
\end{figure}

From the above discussion, we conclude that the scattering does not
occur, in general, between particlelike modes in the tree level of the
interaction. This is an inevitable consequence that follows from the two
assumptions: the $b$-number conservation in Eq.~\eqref{eq2-5-1} and the
number state representation for $b$ particles in Eq.~\eqref{eq2-5-2}.
Momentum transfer does not occur between number states in the tree level
because of the conservation of three momenta.  Schematically, this fact
can be understood by using diagrams shown in Fig.~\ref{fig1}.  In the
usual calculation of S matrix for the scattering process $a+b\to a+b$,
we specify in and out-states as definite particle states for $a$
and $b$ species, but the momentum of each particle can differ between
in and out-states, as shown in Fig.~\ref{fig1} (a).  In the
in-in formalism, this tree level diagram is deformed such that in
and out-states are synchronized [see Fig.~\ref{fig1} (b)].  Then,
the momenta of two in-states must be the same if they are
represented as number states.  It is obvious that there is no momentum
transfer in such a process, and hence the transition is forbidden.  On the other
hand, the momenta of in-states can differ if they are coherent
states, since they are not eigenstates of the number
operator. Therefore, the tree level process is allowed for
self interactions between coherent states, as shown in Fig.~\ref{fig1}
(c).  This is why the second line of Eq.~\eqref{eq2-4-9} has a
nonvanishing contribution.  Figure~\ref{fig1} (d) shows that the
scattering between particlelike modes can occur in the higher order in
perturbation theory.  As will be seen in the Appendix, at least
at the second order in the perturbation theory, this process corresponds to
what we calculate by using the usual Boltzmann equation.

In the above discussion, we assumed that the interaction with other species is given by Eq.~\eqref{eq2-5-1},
which conserves both the axion number and $b$ number.
This can be applied for the gravitational interactions considered in Ref.~\cite{Erken:2011dz}.
However, axions also couple to two photons, 
\begin{equation}
{\cal L}_{a\gamma\gamma} = -\frac{g_{a\gamma\gamma}}{4}\phi F_{\mu\nu}\tilde{F}^{\mu\nu}, \label{eq2-5-6}
\end{equation}
whose number is not conserved.
Here, $F_{\mu\nu}=\partial_{\mu}A_{\nu}-\partial_{\nu}A_{\mu}$ is the photon field strength,
$\tilde{F}^{\mu\nu}=\frac{1}{2}\epsilon^{\mu\nu\lambda\sigma}F_{\lambda\sigma}$ is the dual of it,
$\epsilon^{\mu\nu\lambda\sigma}$ is the totally antisymmetric tensor with $\epsilon^{0123}=+1$,
$g_{a\gamma\gamma}=(\alpha/2\pi F_a)c_{a\gamma\gamma}$ is the axion-photon coupling,
$\alpha$ is the fine structure constant, and $c_{a\gamma\gamma}$ is a numerical coefficient whose value depends on models.
Let us show that this axion-photon coupling does not have any contribution in the tree level of the interaction.

The mode expansion of the photon field is given by
\begin{equation}
A_{\mu}({\bf x},t) = \frac{1}{V}\sum_{(j,r)}\frac{1}{\sqrt{2\omega_j}}\left[e^{ip_j\cdot x}e^r_{j,\mu}b^r_j+e^{-ip_j\cdot x}e^{r*}_{j,\mu}b^{r\dagger}_j\right], \label{eq2-5-7}
\end{equation}
where $\omega_j \equiv |\bf{p}_j|$, $e^r_{j,\mu}$ are polarization vectors and $r$ labels a basis of them.
The creation and annihilation operators satisfy
\begin{equation}
[b^r_j,b^{r'\dagger}_k] = V\delta_{r,r'}\delta_{j,k}, \quad\mathrm{and}\quad [b^r_j,b^{r'}_k]= [b_j^{r\dagger},b_k^{r'\dagger}] = 0. \label{eq2-5-8}
\end{equation}
Substituting Eqs.~\eqref{eq2-2-2} and \eqref{eq2-5-7} into Eq.~\eqref{eq2-5-6}, we obtain the interaction Hamiltonian of the axion-photon coupling
\begin{equation}
H_{I,\gamma}(t) = \frac{1}{V^3}\sum_{i(j,r)(k,r')}\left[\Lambda_{\gamma 1\ i}^{\ \ \ (j,r)(k,r')}e^{-i\Omega_{1i}^{\ jk}t}a_i^{\dagger}b_j^rb_k^{r'}
 + \Lambda_{\gamma 2\ i(j,r)}^{\ \ \ (k,r')}e^{-i\Omega_{2ij}^{\ k}t}a_i^{\dagger}b_j^{r\dagger}b_k^{r'} + \mathrm{H.c.} \right], \label{eq2-5-9}
\end{equation}
where
\begin{align}
\Lambda_{\gamma 1\ i}^{\ \ \ (j,r)(k,r')} &= -\frac{g_{a\gamma\gamma}}{2}\sqrt{\frac{\omega_j\omega_k}{2E_i}}\hat{p}_k\cdot({\bf e}_j^r\times {\bf e}^{r'}_k)V\delta_{i,j+k}, \label{eq2-5-10}\\
\Lambda_{\gamma 2\ i(j,r)}^{\ \ \ (k,r')} &= -\frac{g_{a\gamma\gamma}}{2}\sqrt{\frac{\omega_j\omega_k}{2E_i}}(\hat{p}_j-\hat{p}_k)\cdot({\bf e}_j^{r*}\times {\bf e}^{r'}_k)V\delta_{i+j,k}, \label{eq2-5-11}
\end{align}
$\Omega_{1i}^{\ jk}=\omega_j+\omega_k-E_i$, $\Omega_{2ij}^{\ k}=\omega_k-E_i-\omega_j$, and $\hat{p}_j \equiv {\bf p}_j/|{\bf p}_j|$.
In Eq.~\eqref{eq2-5-9}, we dropped terms containing $a_i b^r_j b^{r'}_k$ and $a_i^{\dagger}b^{r\dagger}_j b^{r'\dagger}_k$,
which violate the conservation of energy in the tree level process.
The commutation relation between the interaction Hamiltonian and the number operator becomes
\begin{equation}
[H_{I,\gamma}(t),\hat{\cal N}_p] = -\frac{1}{V^3}\sum_{(j,r)(k,r')}\left[\Lambda_{\gamma 1\ p}^{\ \ \ (j,r)(k,r')}e^{-i\Omega_{1p}^{\ jk}t}a_p^{\dagger}b_j^r b_k^{r'}
 + \Lambda_{\gamma 2\ p(j,r)}^{\ \ \ (k,r')}e^{-i\Omega_{2pj}^{\ k}t}a_p^{\dagger}b_j^{r\dagger}b_k^{r'} - \mathrm{H.c.} \right]. \label{eq2-5-12}
\end{equation}
Then, it is straightforward to show that the expectation value of Eq.~\eqref{eq2-5-12} vanishes for the particlelike modes $p>K$.
For the condensed modes $p\le K$, we obtain
\begin{equation}
\langle [H_{I,\gamma}(t),\hat{\cal N}_p]\rangle = -\frac{1}{V^{3/2}}\sum_{(j,r)}\left[\Lambda_{\gamma 2\ p(j,r)}^{\ \ \ (j,r)}e^{iE_p t}{\cal N}_{\gamma,j}\alpha_p^*-\mathrm{c.c.} \right]
\quad \mathrm{for} \quad p\le K. \label{eq2-5-13}
\end{equation}
This term also vanishes since $\Lambda_{\gamma 2\ p(j,r)}^{\ \ \ (j,r)}=0$.

The absence of the tree level contribution from the axion-photon interaction can also be understood in terms of some conservation principles.
First of all, the vertex given by $\Lambda_{\gamma 1\ i}^{\ \ \ (j,r)(k,r')}$ violates the number of photons.
Such a process is forbidden at the tree level of the interaction, as long as photons are represented as number states.
On the other hand, the vertex given by $\Lambda_{\gamma 2\ i(j,r)}^{\ \ \ (k,r')}$ conserves the number of photons, leading to Eq.~\eqref{eq2-5-13}.
However, there is no momentum transfer at such a vertex since the momenta of two in-states for photons must be same.
Note that it is possible to show that the tree level contribution vanishes for $p\le K$, without using the structure of $\Lambda_{\gamma 2\ i(j,r)}^{\ \ \ (k,r')}$.
The conservation of three momenta requires that $p=0$ in Eq.~\eqref{eq2-5-13}, hence it is proportional to $e^{im t}$.
The factor $e^{im t}$ gives a rapidly oscillating contribution which can be dropped, since we are interested in the time scale much longer than $\sim m^{-1}$.
Therefore, the tree level process between axions and photons is forbidden due to the conservation of the photon number and three momenta.
\section{\label{sec3}Formation of axion Bose-Einstein condensation}

In this section, we discuss the cosmological evolution of dark matter
axions based on the results in the previous section. The zero mode
axions are produced at the time $t_1$ satisfying the condition
\begin{equation}
m(T_1) = 3H(t_1), \label{eq3-1}
\end{equation}
where $H(t_1)$ is the Hubble parameter at $t_1$ and $T_1$ is the
temperature of radiations at that time. The temperature dependence of
axion mass is obtained in~\cite{Wantz:2009mi},
\begin{equation}
m^2(T) = 1.68\times 10^{-7}\frac{\Lambda_{\mathrm{QCD}}^4}{F_a^2}\left(\frac{T}{\Lambda_{\mathrm{QCD}}}\right)^{-n}, \label{eq3-2}
\end{equation}
with $n=6.68$ and $\Lambda_{\mathrm{QCD}}=400\mathrm{MeV}$.  The time
$t_1$ is estimated as~\cite{Hiramatsu:2012gg}
\begin{equation}
t_1=3.01\times 10^{-7}\mathrm{sec}\left(\frac{g_{*,1}}{70}\right)^{-n/2(4+n)}\left(\frac{F_a}{10^{12}\mathrm{GeV}}\right)^{4/(4+n)}\left(\frac{\Lambda_{\mathrm{QCD}}}{400\mathrm{MeV}}\right)^{-2}, \label{eq3-3}
\end{equation}
where $g_{*,1}$ is the radiation degrees of freedom at the time $t_1$.
This time scale $t_1$ should be identified with $t_0$, which was used in
the previous section.  From Eq.~\eqref{eq2-3-19}, the number density of
the zero mode axions is estimated as
\begin{align}
n(t) &= \frac{1}{2}m(t_1)F_a^2X\left(\frac{R(t_1)}{R(t)}\right)^3 \nonumber\\
&\simeq 2.14\times 10^{47}\mathrm{cm}^{-3}X
\left(\frac{g_{*,1}}{70}\right)^{n/2(4+n)}\left(\frac{F_a}{10^{12}\mathrm{GeV}}\right)^{(4+2n)/(4+n)}\left(\frac{\Lambda_{\mathrm{QCD}}}{400\mathrm{MeV}}\right)^2\left(\frac{R(t_1)}{R(t)}\right)^3, \label{eq3-4}
\end{align}
where $X$ is a numerical factor determined by the initial misalignment
angle $(\theta^{\mathrm{ini}})^2$ and $R(t)$ is the scale factor of the
Universe at the time $t$. Since the momentum dispersion of axions is
given by the horizon at the QCD phase transition $\delta
p(t_1)\sim1/t_1$, their velocity dispersion is estimated as
\begin{equation}
\delta v(t) \sim \frac{\delta p(t)}{m(0)}\sim \frac{1}{m(0)t_1}\left(\frac{R(t_1)}{R(t)}\right) \simeq 3.58\times 10^{-4}
\left(\frac{g_{*,1}}{70}\right)^{n/2(4+n)}\left(\frac{F_a}{10^{12}\mathrm{GeV}}\right)^{n/(4+n)}\left(\frac{R(t_1)}{R(t)}\right), \label{eq3-5}
\end{equation}
where we have used the expression for the zero-temperature axion mass
given by
\begin{equation}
m^2(0) = 1.46\times 10^{-3}\frac{\Lambda_{\mathrm{QCD}}^4}{F_a^2}. \label{eq3-6}
\end{equation}
One can easily verify that the state occupation number of the zero mode
axions is huge,
\begin{equation}
{\cal N} \sim n\frac{(2\pi)^3}{\frac{4\pi}{3}(m\delta v)^3} \sim 10^{61}X\left(\frac{g_{*,1}}{70}\right)^{-n/(4+n)}\left(\frac{F_a}{10^{12}\mathrm{GeV}}\right)^{2(8+n)/(4+n)}\left(\frac{\Lambda_{\mathrm{QCD}}}{400\mathrm{MeV}}\right)^{-4}. \label{eq3-7}
\end{equation}

As was assumed in Sec.~\ref{sec2-3}, $K$ states around the ground state are occupied by ${\cal N}$ particles.
Therefore, each state is occupied by ${\cal N}/K$ particles on average.
Their thermalization rate is given by the time scale of the change of the occupation number for condensed modes~\cite{Erken:2011dz},
\begin{equation}
\Gamma \equiv \frac{1}{{\cal N}_p(t)}\frac{d{\cal N}_p(t)}{dt}, \label{eq3-8}
\end{equation}
where ${\cal N}_p(t)\equiv\langle\hat{{\cal N}}_p(t)\rangle$ is given in
Eq.~\eqref{eq2-4-1} and can be estimated by using the formalism
described in the previous section. Axions form a BEC if this
thermalization rate exceeds the expansion rate
$H(t)$~\cite{Sikivie:2009qn}. Here, it should be kept in mind whether
the system is in the condensed regime or in the particle kinetic
regime. For this purpose, it is necessary to compare $\Gamma$ with the
typical energy dispersion of axions,
\begin{equation}
\delta\omega \sim \frac{1}{2}m(0)(\delta v(t))^2 \sim 
 3.92\times 10^{-13} \mathrm{eV}\left(\frac{g_{*,1}}{70}\right)^{n/(4+n)}\left(\frac{F_a}{10^{12}\mathrm{GeV}}\right)^{-(4-n)/(4+n)}\left(\frac{\Lambda_{\mathrm{QCD}}}{400\mathrm{MeV}}\right)^2\left(\frac{R(t_1)}{R(t)}\right)^2.\label{eq3-9}
\end{equation}

Let us estimate the thermalization rate in the condensed regime.
Substituting the time derivative of Eq.~\eqref{eq2-4-11} into
Eq.~\eqref{eq3-8} yields
\begin{equation}
\Gamma_{\mathrm{condensed}} \simeq \frac{1}{{\cal N}_pV^2}\sum_{j,k,l\le K} \mathrm{Im}\left[\Lambda^{kl}_{pj}\alpha_k\alpha_l\alpha_j^*\alpha_p^*\right], \label{eq3-10}
\end{equation}
where we have used the approximation~\eqref{eq2-4-13}. Defining the
factor $\Lambda$ such that
\begin{equation}
\Lambda^{kl}_{pj} = \Lambda V\delta_{k+l,p+j}, \label{eq3-11}
\end{equation}
and using the approximation ${\cal N}_p \simeq |\alpha_p|^2\simeq {\cal N}/K$,
we finally obtain
\begin{equation}
\Gamma_{\mathrm{condensed}} \simeq \Lambda\frac{{\cal N}}{V} = \Lambda n(t), \label{eq3-12}
\end{equation}
where $n(t)$ is the number density of the zero mode axions given in Eq.~\eqref{eq3-4}.
For a $\lambda\phi^4$ type self interaction, the
expression of $\Lambda$ in Eq.~\eqref{eq2-4-3} gives
\begin{equation}
\Gamma_{\mathrm{condensed},s} \simeq \frac{\lambda n(t)}{4m^2}. \label{eq3-13} 
\end{equation}
On the other hand, the expression of $\Lambda$ in Eq.~\eqref{eq2-4-5}
for gravitational self interaction leads to
\begin{equation}
\Gamma_{\mathrm{condensed},g} \simeq \frac{4\pi Gm^2n(t)}{(\delta p(t))^2}, \label{eq3-14}
\end{equation}
where $\delta p(t)$ is the momentum dispersion of axions [see
Eq.~\eqref{eq3-5}].  Note that these expressions are valid only if the
condition $\delta\omega\ll \Gamma_{\mathrm{condensed}}$ is satisfied.

In the opposite case $\delta \omega \gg \Gamma$, the expression in the
particle kinetic regime must be used. As seen in Sec.~\ref{sec2-4}, the
first order term vanishes in this regime, which requires us to evaluate
the second order terms in order to estimate the transition rate of the
occupation number. Since the thermalization rate is the quantity of
${\cal O}(\Lambda^2)$, it is suppressed compared to that in the
condensed regime by a factor of $\Lambda$. Thus, one can expect that it
is difficult for axions to thermalize in this particle kinetic
regime. In fact, it is also obvious from the fact that the conditions
$\Gamma_{\mathrm{particle}}>H$ (thermalization condition) and
$\Gamma_{\mathrm{particle}} < \delta \omega$ (particle kinetic
condition) are incompatible due to $\delta \omega < H$ after the time
$t_1$ [see Eqs.~\eqref{eq3-5} and \eqref{eq3-9}]. Here,
$\Gamma_{\mathrm{particle}}$ is the transition rate obtained from the
second order terms in the perturbative calculation. Thus, we can
conclude that the axion thermalization occurs only in the condensed
regime.

\begin{figure}[htbp]
\begin{center}
\includegraphics[scale=1.0,angle=0]{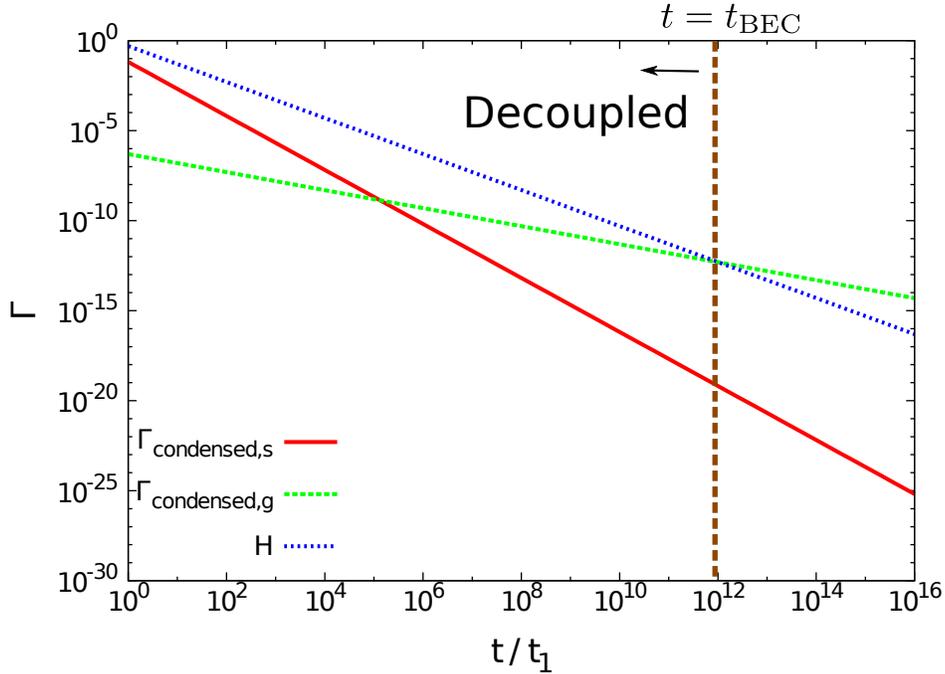}
\end{center}
\caption{The time evolution of the relaxation rates $\Gamma$ in the
condensed regime and the expansion rate $H$.
$\Gamma_{\mathrm{condensed},s}$ is the relaxation rate due to
$\lambda\phi^4$ interaction estimated in Eq.~\eqref{eq3-13} and
$\Gamma_{\mathrm{condensed},g}$ is the relaxation rate due to
gravitational self interaction estimated in Eq.~\eqref{eq3-14}.  In this
figure we normalize all the scales in unit of $t_1=1$, where $t_1$ is
given in Eq.~\eqref{eq3-3}, and the following parameter values, $X=1$,
$g_{*,1}=70$, and $F_a=10^{12}\mathrm{GeV}$, are taken. Axions
begin to develop toward a BEC
at the time $t_{\mathrm{BEC}}$ given in Eq.~\eqref{eq3-16}.}
\label{fig2}
\end{figure}

Figure~\ref{fig2} shows the time evolution of thermalization rates
$\Gamma$ together with the expansion rate $H$.  We find that the
transition rate $\Gamma_{\mathrm{condensed},g}$ due to the gravitational
self interaction exceeds the expansion rate when the temperature of
photons becomes
\begin{equation}
T_{\mathrm{BEC}} \simeq 2.07\times 10^3 
\mathrm{eV}\ X\left(\frac{g_{*,1}}{70}\right)^{-3n/4(4+n)}\left(\frac{F_a}{10^{12}\mathrm{GeV}}\right)^{6/(4+n)}\left(\frac{\Lambda_{\mathrm{QCD}}}{400\mathrm{MeV}}\right), 
\label{eq3-15}
\end{equation}
which corresponds to the time scale
\begin{equation}
t_{\mathrm{BEC}} \simeq 3.09\times 10^5\mathrm{sec}\ X^{-2}\left(\frac{g_{*,1}}{70}\right)^{3n/2(4+n)}\left(\frac{F_a}{10^{12}\mathrm{GeV}}\right)^{-12/(4+n)}\left(\frac{\Lambda_{\mathrm{QCD}}}{400\mathrm{MeV}}\right)^{-2}.  
\label{eq3-16}
\end{equation}

Before the time $t_{\mathrm{BEC}}$, axions are decoupled from each other
and are described as a classical field.  Once a BEC is formed, however,
axions behave like cold dark matter for a different reason from that of the
classical field. In particular, from the causality, the correlation
length $l$ of the axion field is expected to extend over the horizon
$l\lesssim t$~\cite{Erken:2011dz}.  Hence the momentum dispersion
$\delta p$ appearing in Eq.~\eqref{eq3-14} becomes comparable with
$l^{-1}\sim t^{-1}$, which makes the time scale of the thermalization
process much faster. Then, axions continue to rethermalize themselves,
and almost all axions stay in the lowest energy state.  This leads
to the modifications of some quantities such as the energy-momentum
tensor and the evolution equation of density perturbations, but they do
not induce any effect on the length scale relevant to cosmological
observations~\cite{Sikivie:2009qn,Hwang:2009js}.

It should be emphasized that we have not completely proven the formation of a BEC.
In order to confirm that axions do form a BEC, it is necessary to show that they establish the Bose-Einstein distribution after the time $t_{\mathrm{BEC}}$.
This requires us to investigate the evolution of the distribution function in a more precise way.
The formation of a BEC far from equilibrium states is investigated in~\cite{Berges:2012us}.
According to the result of~\cite{Berges:2012us}, at first the system enters into a universal distribution described by the power law, 
and subsequently it redistributes into the Bose-Einstein distribution.
For dark matter axions, we expect that the universal distribution appears soon after the time $t_{\mathrm{BEC}}$.
However, it is not so obvious that it redistributes into the Bose-Einstein form within the time scale $t_{\mathrm{BEC}}$.
In order to establish the equilibrium distribution, the system must create some particlelike modes from condensed modes, since these particlelike modes
should appear as the higher momentum states in the Bose-Einstein distribution.
Such an evolution toward the equilibrium regime cannot be calculated in the formalism developed in this paper.
Therefore, although the result of our analysis indicates that the zero mode axions begin to evolve toward thermal equilibrium at $t=t_{\mathrm{BEC}}$,
the question of the establishment of the Bose-Einstein distribution is still open.
It is necessary to develop the computational method to calculate the evolution of the distribution function from out of equilibrium to equilibrium,
which is out of the scope of this work.

In Refs.~\cite{Erken:2011vv,Erken:2011dz}, it was pointed out that these
excited modes would be relativistic and affect the cosmological
parameters such as the baryon to photon ratio and the effective number
of neutrino species, if axions enter into thermal contact with
photons. However, as shown in Sec.~\ref{sec2-5}, the interaction between
axions and other species exactly vanishes at the first order in
the perturbation theory
if we assume that other species are represented as number states,
and hence the thermalization rate with other
species is heavily suppressed, which indicates that axions forming a BEC
are decoupled from other particles and do not give any significant
modifications to cosmological parameters.  Even though, if an axion BEC is
the dominant component of dark matter, they give some imprints on the
structure of the inner caustics of galactic halos~\cite{Sikivie:2010bq}.
This might be a useful tool to distinguish axion dark matter from other
particle dark matter candidates.

\section{\label{sec4}Summary and conclusion}
In this paper, we developed the formalism to describe the evolution of
the zero mode axions in terms of the quantum field theoretic method.  In
order to solve the evolution of occupation number of axions, we used the
in-in formalism and the coherent state representation for a highly
degenerate Bose gas of axions. Combining these two ingredients, we
derive the time evolution of the expectation value of the number
operator in a perturbative way. We showed that there is a nonvanishing
contribution for the self interaction of condensed modes at the leading
order in the perturbation theory [see Eq.~\eqref{eq2-4-11}]. On the other
hand, the interactions between particlelike modes including other
species exactly vanish at the first order in perturbative expansion,
which indicates that their relaxation rate is suppressed.

Using the results for the time evolution of the occupation number, we
estimated the thermalization rate of the zero mode axions.  We recovered
the expressions for thermalization rates obtained
in~\cite{Erken:2011dz}, and confirmed that axions 
start to develop toward a BEC due to the
gravitational self interactions when the temperature of photons becomes
${\cal O}(10^3)\mathrm{eV}$.

From the fact that the tree level contribution vanishes for the
interaction with other particlelike species, we conclude that axions
in the condensed regime
have no thermal contact with other cosmological fluids, and that
there is no significant effect on cosmological parameters.  In
particular, for the effective number of neutrino species
$N_{\mathrm{eff}}$, the result of~\cite{Erken:2011vv,Erken:2011dz}
predicts the higher value $N_{\mathrm{eff}}=6.77$ than the standard
model, but in our analysis $N_{\mathrm{eff}}$ does not differ from the
standard value.  Hence the axion BEC is consistent with the standard
cosmology. The only peculiar prediction is the specific phase space
structure for galactic halos~\cite{Sikivie:2010bq}, which gives a
possibility to probe axion BEC dark matter on observational grounds.

Finally, let us comment on speculative points in our discussion.
For the gravitational self interaction of axions, we use the expression~\eqref{eq2-4-4} which holds in the Newtonian limit.
The use of this term is justified only if we are able to ignore the requirement of the causality.
To be more precise, Eq.~\eqref{eq2-4-4} can be regarded as a good approximation while the time scale of the interaction $\Gamma^{-1}$
exceeds the typical length scale $\delta l\sim \delta p^{-1}$ on which the interaction takes place.
However, it is expected that
$\delta p\lesssim H$ for condensed modes,
and hence this condition becomes $\Gamma^{-1}>H^{-1}$,
which seems to be incompatible with the thermalization condition $\Gamma> H$.
Therefore, in order to make a clear description about the rethermalization of the axion BEC, we must extend our formalism
into the expanding background, including the correction coming from general relativity.

In the present analysis, we just estimate the time scale at which the self interaction of condensed modes becomes relevant, but
the completion of BEC requires the occurrence of the transition from condensed modes to particlelike modes.
This effect is not included in the formalism developed in this work.
Furthermore, even if the formation of the BEC completes within the time scale estimated in $t_{\mathrm{BEC}}$,
there remains a question whether this rethermalizing BEC is disturbed when the nonlinear structure grows.
It is important to confirm the rethermalization property of dark matter axions infalling to galactic halos,
since this property is claimed to explain the caustic rings, giving a motivation of axion BEC dark matter~\cite{Sikivie:2010bq}.
Each of the issues enumerated above requires more extensive studies, which should be addressed
in future publications.
\begin{acknowledgments}
We would like to thank Pierre Sikivie for his useful comments.
K.~S.~is supported by the Japan Society for the Promotion of Science (JSPS)
through research fellowships. M.~Y.~is supported in part by the
Grant-in-Aid for Scientific Research No.~21740187 and the Grant-in-Aid
for Scientific Research on Innovative Areas No.~24111706.
\end{acknowledgments}

\appendix

\section{Second order in perturbation theory}
\label{secA}
In this appendix, we compute the terms of second order in $H_I$ [the
third term of the right-hand side of Eq.~\eqref{eq2-4-1}].
In the second order contributions, it is possible to include axion number violating vertices, which was dropped in Eq.~\eqref{eq2-4-2}.
Here we omit such terms just for simplicity.
Inclusion of axion number violating interactions might modify the results, but it would not affect the main line of derivations.
From Eqs.~\eqref{eq2-4-2} and \eqref{eq2-4-7}, we obtain
\begin{equation}
[H_I(t),{\cal \hat{N}}_p] = \frac{1}{2V^4}\sum_{jkl}\left[\Lambda^{pj}_{kl}e^{-i\Omega^{pj}_{kl}t}a^{\dagger}_k a^{\dagger}_l a_ja_p - \mathrm{H.c.}\right]. \label{eqA-1}
\end{equation}
Then we find
\begin{eqnarray}
[H_I(t_1),[H_I(t_2),{\cal \hat{N}}_p]]
&=& \frac{1}{8}\frac{1}{V^8}\sum_{mnqrjkl}\Lambda^{mn}_{qr}\Lambda^{pj}_{kl}e^{-i(\Omega^{mn}_{qr}t_1+\Omega^{pj}_{kl}t_2)}[a_q^{\dagger}a_r^{\dagger}a_ma_n,a_k^{\dagger}a_l^{\dagger}a_ja_p] \nonumber\\
&&-\frac{1}{8}\frac{1}{V^8}\sum_{mnqrjkl}\Lambda^{mn}_{qr}\Lambda^{kl}_{pj}e^{-i(\Omega^{mn}_{qr}t_1-\Omega^{pj}_{kl}t_2)}[a_q^{\dagger}a_r^{\dagger}a_ma_n,a_p^{\dagger}a_j^{\dagger}a_ka_l]. \label{eqA-2}
\end{eqnarray}
Note that
\begin{eqnarray}
[a_q^{\dagger}a_r^{\dagger}a_ma_n,a_k^{\dagger}a_l^{\dagger}a_ja_p] &=&
V^2\left[(\delta_{nl}\delta_{mk}+\delta_{nk}\delta_{ml})a_q^{\dagger}a_r^{\dagger}a_ja_p - (\delta_{jq}\delta_{pr}+\delta_{pq}\delta_{jr})a_k^{\dagger}a_l^{\dagger}a_ma_n \right]\nonumber\\
&&+V\left[\delta_{ml}a_q^{\dagger}a_k^{\dagger}a_r^{\dagger}a_ja_pa_n + \delta_{mk}a_q^{\dagger}a_l^{\dagger}a_r^{\dagger}a_ja_pa_n
- \delta_{jq}a_k^{\dagger}a_l^{\dagger}a_r^{\dagger}a_pa_ma_n - \delta_{pq}a_k^{\dagger}a_l^{\dagger}a_r^{\dagger}a_ja_ma_n \right. \nonumber\\
&&\left. +\delta_{nl}a_q^{\dagger}a_r^{\dagger}a_k^{\dagger}a_ma_ja_p + \delta_{nk}a_q^{\dagger}a_r^{\dagger}a_l^{\dagger}a_ma_ja_p
- \delta_{jr}a_q^{\dagger}a_k^{\dagger}a_l^{\dagger}a_ma_pa_n - \delta_{pr}a_q^{\dagger}a_k^{\dagger}a_l^{\dagger}a_ma_ja_n \right]. \label{eqA-3}
\end{eqnarray}
Using this formula, after some simplification, we reduce the expectation value of
Eq.~\eqref{eqA-2} into
\begin{eqnarray}
\langle[H_I(t_1),[H_I(t_2),{\cal \hat{N}}_p]]\rangle
&=& \frac{1}{4}\frac{1}{V^6}\sum_{qjklm}\left[\Lambda^{qm}_{kl}\Lambda^{pj}_{qm}e^{-i(\Omega^{qm}_{kl}t_1+\Omega^{pj}_{qm}t_2)}\langle a_k^{\dagger}a_l^{\dagger}a_ja_p\rangle+\mathrm{c.c.}\right] \nonumber\\
&&-\frac{1}{4}\frac{1}{V^6}\sum_{qjklm}\left[\Lambda^{jq}_{pm}\Lambda^{pm}_{kl}e^{-i(\Omega^{qj}_{pm}t_1+\Omega^{pm}_{kl}t_2)}\langle a_k^{\dagger}a_l^{\dagger}a_ja_q\rangle+\mathrm{c.c.}\right] \nonumber\\
&&+\frac{1}{2}\frac{1}{V^7}\sum_{nmqjkl}\left[\Lambda^{ln}_{qm}\Lambda^{pj}_{kl}e^{-i(\Omega^{ln}_{qm}t_1+\Omega^{pj}_{kl}t_2)}\langle a_q^{\dagger}a_m^{\dagger}a_k^{\dagger}a_pa_ja_n\rangle+\mathrm{c.c.}\right] \nonumber\\
&&-\frac{1}{4}\frac{1}{V^7}\sum_{nmqjkl}\left[\Lambda^{mn}_{ql}\Lambda^{pl}_{kj}e^{-i(\Omega^{mn}_{ql}t_1+\Omega^{pl}_{kj}t_2)}\langle a_k^{\dagger}a_j^{\dagger}a_q^{\dagger}a_pa_ma_n\rangle+\mathrm{c.c.}\right] \nonumber\\
&&-\frac{1}{4}\frac{1}{V^7}\sum_{nmqjkl}\left[\Lambda^{mn}_{pq}\Lambda^{pj}_{kl}e^{-i(\Omega^{mn}_{pq}t_1+\Omega^{pj}_{kl}t_2)}\langle a_k^{\dagger}a_l^{\dagger}a_q^{\dagger}a_ja_ma_n\rangle+\mathrm{c.c.}\right]. \label{eqA-4}
\end{eqnarray}

To compute the expectation value, we note that
\begin{equation}
a_{k''}a_{k'}a_k |\{{\cal N}\},\{\alpha\}\rangle =
\left\{
\renewcommand{\arraystretch}{1.8}
\begin{array}{l l l}
\sqrt{{\cal N}_k({\cal N}_k-1)({\cal N}_k-2)}V^{3/2}|\{{\cal N}\}^{3k}, \{\alpha\} \rangle & \mathrm{if} & k=k'=k''>K \\
\sqrt{{\cal N}_k{\cal N}_{k'}({\cal N}_{k'}-1)}V^{3/2}|\{{\cal N}\}^{k,2k'}, \{\alpha\} \rangle & \mathrm{if} & k\ne k',\ k'=k''>K,\ \mathrm{and}\ k>K \\
\sqrt{{\cal N}_k{\cal N}_{k'}{\cal N}_{k''}}V^{3/2}|\{{\cal N}\}^{k,k',k''}, \{\alpha\} \rangle & \mathrm{if} & k\ne k'\ne k'',\ k>K,\ k'>K,\ \mathrm{and}\ k''>K\\
\sqrt{{\cal N}_k({\cal N}_k-1)}\alpha_{k''}V^{3/2}|\{{\cal N}\}^{2k}, \{\alpha\} \rangle & \mathrm{if} & k=k'>K\ \mathrm{and}\ k''\le K \\
\sqrt{{\cal N}_k{\cal N}_{k'}}\alpha_{k''}V^{3/2}|\{{\cal N}\}^{kk'}, \{\alpha\} \rangle & \mathrm{if} & k\ne k',\ k>K,\ k'>K,\ \mathrm{and}\ k''\le K \\
\sqrt{{\cal N}_k}\alpha_{k'}\alpha_{k''}V^{3/2}|\{{\cal N}\}^k, \{\alpha\} \rangle & \mathrm{if} & k>K,\ k'\le K,\ \mathrm{and}\ k''\le K \\
\alpha_k\alpha_{k'}\alpha_{k''}V^{3/2}|\{{\cal N}\}, \{\alpha\} \rangle & \mathrm{if} & k\le K,\ k'\le K,\ \mathrm{and}\ k''\le K \\
\end{array}
\renewcommand{\arraystretch}{1}
\right., \label{eqA-5}
\end{equation}
where the state $|\{{\cal N}\}^{3k}, \{\alpha\} \rangle$ contains a
factor $(a^{\dagger}_k)^{{\cal N}_k-3}/\sqrt{({\cal N}_k-3)!V^{{\cal
N}_k-3}}$ for mode $k$, the state $|\{{\cal N}\}^{k,2k'}, \{\alpha\}
\rangle$ contains a factor $(a^{\dagger}_k)^{{\cal
N}_k-1}(a^{\dagger}_{k'})^{{\cal N}_{k'}-2}/\sqrt{({\cal
N}_k-1)!V^{{\cal N}_k-1}({\cal N}_{k'}-2)!V^{{\cal N}_{k'}-2}}$ for
modes $k$ and $k'$, and the state $|\{{\cal N}\}^{k,k',k''}, \{\alpha\}
\rangle$ contains a factor $(a^{\dagger}_k)^{{\cal
N}_k-1}(a^{\dagger}_{k'})^{{\cal N}_{k'}-1}(a^{\dagger}_{k''})^{{\cal
N}_{k''}-1}/\sqrt{({\cal N}_k-1)!V^{{\cal N}_k-1}({\cal
N}_{k'}-1)!V^{{\cal N}_{k'}-1}({\cal N}_{k''}-1)!V^{{\cal N}_{k''}-1}}$
for modes $k$, $k'$ and $k''$.  Using Eq.~\eqref{eqA-5}, the expectation
value of Eq.~\eqref{eqA-4} can be evaluated in a similar way to the
first order terms.  With a tedious but straightforward calculation, the
first line of Eq.~\eqref{eqA-4} becomes
\begin{eqnarray}
 &&\frac{1}{4}\frac{1}{V^6}\sum_{qjklm}\left[\Lambda^{qm}_{kl}\Lambda^{pj}_{qm}e^{-i(\Omega^{qm}_{kl}t_1+\Omega^{pj}_{qm}t_2)}\langle a_k^{\dagger}a_l^{\dagger}a_ja_p\rangle+\mathrm{c.c.}\right] \nonumber\\
 &&\quad= \frac{1}{2}\frac{1}{V^4}\sum_{qm}\sum_{k\le K}\sum_{l>K}\left[\Lambda^{qm}_{kl}\Lambda^{pl}_{qm}e^{-i(\Omega^{qm}_{kl}t_1+\Omega^{pl}_{qm}t_2)}\alpha^*_k\alpha_p{\cal N}_l + \mathrm{c.c.} \right] \nonumber\\
 &&\qquad+  \frac{1}{4}\frac{1}{V^4}\sum_{qm}\sum_{j,k,l\le K}\left[\Lambda^{qm}_{kl}\Lambda^{pj}_{qm}e^{-i(\Omega^{qm}_{kl}t_1+\Omega^{pj}_{qm}t_2)}\alpha^*_k\alpha^*_l\alpha_j\alpha_p + \mathrm{c.c.} \right] 
 \quad \mathrm{for}\quad p\le K, \label{eqA-6}
 \end{eqnarray}
or
\begin{eqnarray}
 &&\frac{1}{4}\frac{1}{V^6}\sum_{qjklm}\left[\Lambda^{qm}_{kl}\Lambda^{pj}_{qm}e^{-i(\Omega^{qm}_{kl}t_1+\Omega^{pj}_{qm}t_2)}\langle a_k^{\dagger}a_l^{\dagger}a_ja_p\rangle+\mathrm{c.c.}\right] \nonumber\\
 &&\quad= \frac{1}{4}\frac{1}{V^4}\sum_{qm}\left[\Lambda^{qm}_{pp}\Lambda^{pp}_{qm}e^{-i(\Omega^{qm}_{pp}t_1+\Omega^{pp}_{qm}t_2)}{\cal N}_p({\cal N}_p-1) + \mathrm{c.c.} \right] \nonumber\\
 &&\qquad+  \frac{1}{2}\frac{1}{V^4}\sum_{qm}\sum_{j>K,j\ne p}|\Lambda^{qm}_{pj}|^2\left[e^{-i\Omega^{qm}_{pj}(t_1-t_2)}{\cal N}_p{\cal N}_j + \mathrm{c.c.} \right] \nonumber\\
 &&\qquad+  \frac{1}{2}\frac{1}{V^4}\sum_{qm}\sum_{j,k\le K}\left[\Lambda^{qm}_{pk}\Lambda^{pj}_{qm}e^{-i(\Omega^{qm}_{pk}t_1+\Omega^{pj}_{qm}t_2)}\alpha^*_k\alpha_j{\cal N}_p + \mathrm{c.c.} \right]  
 \quad \mathrm{for}\quad p> K. \label{eqA-7}
 \end{eqnarray}
The second line of Eq.~\eqref{eqA-4} becomes 
\begin{eqnarray}
&&-\frac{1}{4}\frac{1}{V^6}\sum_{qjklm}\left[\Lambda^{jq}_{pm}\Lambda^{pm}_{kl}e^{-i(\Omega^{qj}_{pm}t_1+\Omega^{pm}_{kl}t_2)}\langle a_k^{\dagger}a_l^{\dagger}a_ja_q\rangle+\mathrm{c.c.}\right] \nonumber \\
&&\quad = -\frac{1}{4}\frac{1}{V^4}\sum_{m}\sum_{q>K}\left[\Lambda^{qq}_{pm}\Lambda^{pm}_{qq}e^{-i(\Omega^{qq}_{pm}t_1+\Omega^{pm}_{qq}t_2)}{\cal N}_q({\cal N}_q-1) + \mathrm{c.c.} \right] \nonumber\\
 &&\qquad-  \frac{1}{2}\frac{1}{V^4}\sum_{m}\sum_{j>K,j\ne p}\sum_{q>K}|\Lambda^{jq}_{pm}|^2\left[e^{-i\Omega^{qj}_{pm}(t_1-t_2)}{\cal N}_j{\cal N}_q + \mathrm{c.c.} \right] \nonumber\\
 &&\qquad-  \frac{1}{V^4}\sum_{m}\sum_{j,k\le K}\sum_{l>K}\left[\Lambda^{jl}_{pm}\Lambda^{pm}_{kl}e^{-i(\Omega^{jl}_{pm}t_1+\Omega^{pm}_{kl}t_2)}\alpha^*_k\alpha_j{\cal N}_l + \mathrm{c.c.} \right] \nonumber\\
  &&\qquad- \frac{1}{4} \frac{1}{V^4}\sum_{m}\sum_{j,k,l,q\le K}\left[\Lambda^{jq}_{pm}\Lambda^{pm}_{kl}e^{-i(\Omega^{jq}_{pm}t_1+\Omega^{pm}_{kl}t_2)}\alpha^*_k\alpha^*_l\alpha_j\alpha_q + \mathrm{c.c.} \right].
  \label{eqA-8}
\end{eqnarray}
The third line of Eq.~\eqref{eqA-4} becomes
\begin{eqnarray}
&&\frac{1}{2}\frac{1}{V^7}\sum_{nmqjkl}\left[\Lambda^{ln}_{qm}\Lambda^{pj}_{kl}e^{-i(\Omega^{ln}_{qm}t_1+\Omega^{pj}_{kl}t_2)}\langle a_q^{\dagger}a_m^{\dagger}a_k^{\dagger}a_pa_ja_n\rangle+\mathrm{c.c.}\right] \nonumber\\
&&\quad= \frac{1}{V^4}\sum_{l}\sum_{q\le K}\sum_{k>K}\left[\Lambda^{lk}_{qk}\Lambda^{pk}_{kl}e^{-i(\Omega^{lk}_{qk}t_1+\Omega^{pk}_{kl}t_2)}{\cal N}_k({\cal N}_k-1)\alpha^*_q\alpha_p + \mathrm{c.c.} \right] \nonumber\\
&&\qquad + \frac{1}{2}\frac{1}{V^4}\sum_{l}\sum_{q\le K}\sum_{k>K}\left[\Lambda^{lk}_{kk}\Lambda^{pk}_{ql}e^{-i(\Omega^{lk}_{kk}t_1+\Omega^{pk}_{ql}t_2)}{\cal N}_k({\cal N}_k-1)\alpha^*_q\alpha_p + \mathrm{c.c.} \right] \nonumber\\
&& \qquad +\frac{1}{V^4}\sum_{l}\sum_{q\le K}\sum_{m\ne k,m>K}\sum_{k>K}\left[{\cal N}_m{\cal N}_k\alpha_q^*\alpha_p \right. \nonumber \\
&&\qquad\quad \left.\times \left(\Lambda^{lk}_{qm}\Lambda^{pm}_{kl}e^{-i(\Omega^{lk}_{qm}t_1+\Omega^{pm}_{kl}t_2)}
+ \Lambda^{lm}_{qm}\Lambda^{pk}_{kl}e^{-i(\Omega^{lm}_{qm}t_1+\Omega^{pk}_{kl}t_2)} 
+ \Lambda^{lk}_{mk}\Lambda^{pm}_{ql}e^{-i(\Omega^{lk}_{mk}t_1+\Omega^{pm}_{ql}t_2)}\right)
 + \mathrm{c.c.} \right] \nonumber\\
&& \qquad +\frac{1}{V^4}\sum_{l}\sum_{m,n,k\le K}\sum_{q>K}\left[{\cal N}_q\alpha_m^*\alpha_k^*\alpha_p\alpha_n 
 \left(\Lambda^{ln}_{qm}\Lambda^{pq}_{kl}e^{-i(\Omega^{ln}_{qm}t_1+\Omega^{pq}_{kl}t_2)}
+ \Lambda^{lq}_{qm}\Lambda^{pn}_{kl}e^{-i(\Omega^{lq}_{qm}t_1+\Omega^{pn}_{kl}t_2)}\right)
 + \mathrm{c.c.} \right] \nonumber\\
 && \qquad + \frac{1}{2}\frac{1}{V^4}\sum_{l}\sum_{m,n,k\le K}\sum_{q>K}\left[{\cal N}_q\alpha_m^*\alpha_k^*\alpha_p\alpha_n 
 \left(\Lambda^{ln}_{km}\Lambda^{pq}_{ql}e^{-i(\Omega^{ln}_{km}t_1+\Omega^{pq}_{ql}t_2)}
+ \Lambda^{lq}_{km}\Lambda^{pn}_{ql}e^{-i(\Omega^{lq}_{km}t_1+\Omega^{pn}_{ql}t_2)}\right)
 + \mathrm{c.c.} \right] \nonumber\\
 && \qquad + \frac{1}{2}\frac{1}{V^4}\sum_{l}\sum_{q,m,k,j,n\le K}\left[\Lambda^{ln}_{qm}\Lambda^{pj}_{kl}e^{-i(\Omega^{ln}_{qm}t_1+\Omega^{pj}_{kl}t_2)}
 \alpha^*_q\alpha^*_m\alpha^*_k\alpha_p\alpha_j\alpha_n + \mathrm{c.c.} \right] \quad \mathrm{for}\quad p\le K, \label{eqA-9}
\end{eqnarray}
or
\begin{eqnarray}
&&\frac{1}{2}\frac{1}{V^7}\sum_{nmqjkl}\left[\Lambda^{ln}_{qm}\Lambda^{pj}_{kl}e^{-i(\Omega^{ln}_{qm}t_1+\Omega^{pj}_{kl}t_2)}\langle a_q^{\dagger}a_m^{\dagger}a_k^{\dagger}a_pa_ja_n\rangle+\mathrm{c.c.}\right] \nonumber\\
&&\quad= \frac{1}{2}\frac{1}{V^4}\sum_{l}\left[\Lambda^{lp}_{pp}\Lambda^{pp}_{pl}e^{-i(\Omega^{lp}_{pp}t_1+\Omega^{pp}_{pl}t_2)}{\cal N}_p({\cal N}_p-1)({\cal N}_p-2) + \mathrm{c.c.} \right] \nonumber\\
&&\qquad + \frac{1}{V^4}\sum_{l}\sum_{q\ne p,q> K}\left[{\cal N}_p({\cal N}_p-1){\cal N}_q
\left(\Lambda^{lq}_{qp}\Lambda^{pp}_{pl}e^{-i(\Omega^{lq}_{qp}t_1+\Omega^{pp}_{pl}t_2)}
+ \Lambda^{lp}_{qp}\Lambda^{pq}_{pl}e^{-i(\Omega^{lp}_{qp}t_1+\Omega^{pq}_{pl}t_2)}\right) + \mathrm{c.c.} \right] \nonumber\\
&&\qquad + \frac{1}{2}\frac{1}{V^4}\sum_{l}\sum_{q\ne p,q> K}\left[{\cal N}_p({\cal N}_p-1){\cal N}_q
\left(\Lambda^{lq}_{pp}\Lambda^{pp}_{ql}e^{-i(\Omega^{lq}_{pp}t_1+\Omega^{pp}_{ql}t_2)}
+ \Lambda^{lp}_{pp}\Lambda^{pq}_{ql}e^{-i(\Omega^{lp}_{pp}t_1+\Omega^{pq}_{ql}t_2)}\right) + \mathrm{c.c.} \right] \nonumber\\
&&\qquad + \frac{1}{V^4}\sum_{l}\sum_{q\ne p,q>K}\left[\Lambda^{lq}_{pq}\Lambda^{pq}_{ql}e^{-i(\Omega^{lq}_{pq}t_1+\Omega^{pq}_{ql}t_2)}{\cal N}_q({\cal N}_q-1){\cal N}_p + \mathrm{c.c.} \right] \nonumber\\
&&\qquad + \frac{1}{2}\frac{1}{V^4}\sum_{l}\sum_{q\ne p,q>K}\left[\Lambda^{lq}_{qq}\Lambda^{pq}_{pl}e^{-i(\Omega^{lq}_{qq}t_1+\Omega^{pq}_{pl}t_2)}{\cal N}_q({\cal N}_q-1){\cal N}_p + \mathrm{c.c.} \right] \nonumber\\
&&\qquad + \frac{1}{V^4}\sum_{l}\sum_{q\ne p,q>K}\sum_{k\ne q,k\ne p,k>K}\left[|\Lambda^{lk}_{pq}|^2e^{-i\Omega^{lk}_{pq}(t_1-t_2)}{\cal N}_p{\cal N}_q{\cal N}_k + \mathrm{c.c.} \right] \nonumber\\
&&\qquad + \frac{1}{V^4}\sum_{l}\sum_{q\ne p,q>K}\sum_{k\ne q,k\ne p,k>K}\left[\Lambda^{lk}_{qk}\Lambda^{pq}_{pl}e^{-i(\Omega^{lk}_{qk}t_1+\Omega^{pq}_{pl}t_2)}{\cal N}_p{\cal N}_q{\cal N}_k + \mathrm{c.c.} \right] \nonumber\\
&&\qquad +\frac{1}{V^4}\sum_{l}\sum_{q\ne p,q>K}\sum_{k\ne q,k\ne p,k>K}\left[\Lambda^{lk}_{pk}\Lambda^{pq}_{ql}e^{-i(\Omega^{lk}_{pk}t_1+\Omega^{pq}_{ql}t_2)}{\cal N}_p{\cal N}_q{\cal N}_k + \mathrm{c.c.} \right] \nonumber\\
&&\qquad +\frac{1}{V^4}\sum_{l}\sum_{n,m\le K}\left[{\cal N}_p({\cal N}_p-1)\alpha_n^*\alpha_m
\left(\Lambda^{lp}_{np}\Lambda^{pm}_{pl}e^{-i(\Omega^{lp}_{np}t_1+\Omega^{pm}_{pl}t_2)}
+ \Lambda^{lm}_{np}\Lambda^{pp}_{pl}e^{-i(\Omega^{lm}_{np}t_1+\Omega^{pp}_{pl}t_2)}
\right) + \mathrm{c.c.} \right] \nonumber\\
&&\qquad +\frac{1}{2}\frac{1}{V^4}\sum_{l}\sum_{n,m\le K}\left[{\cal N}_p({\cal N}_p-1)\alpha_n^*\alpha_m
\left(\Lambda^{lp}_{pp}\Lambda^{pm}_{nl}e^{-i(\Omega^{lp}_{pp}t_1+\Omega^{pm}_{nl}t_2)}
+ \Lambda^{lm}_{pp}\Lambda^{pp}_{nl}e^{-i(\Omega^{lm}_{pp}t_1+\Omega^{pp}_{nl}t_2)}
\right) + \mathrm{c.c.} \right] \nonumber
\end{eqnarray}
\begin{eqnarray}
&&\qquad +\frac{1}{V^4}\sum_{l}\sum_{n,m\le K}\sum_{q\ne p,q>K}\left[{\cal N}_p{\cal N}_q\alpha_n^*\alpha_m  \right.\nonumber\\
&&\qquad\quad \times\left(\Lambda^{lq}_{np}\Lambda^{pm}_{ql}e^{-i(\Omega^{lq}_{np}t_1+\Omega^{pm}_{ql}t_2)}
+ \Lambda^{lq}_{nq}\Lambda^{pm}_{pl}e^{-i(\Omega^{lq}_{nq}t_1+\Omega^{pm}_{pl}t_2)}
+\Lambda^{lm}_{np}\Lambda^{pq}_{ql}e^{-i(\Omega^{lm}_{np}t_1+\Omega^{pq}_{ql}t_2)} \right.\nonumber\\
&&\qquad\qquad \left.\left.+ \Lambda^{lm}_{nq}\Lambda^{pq}_{pl}e^{-i(\Omega^{lm}_{nq}t_1+\Omega^{pq}_{pl}t_2)}
+ \Lambda^{lq}_{pq}\Lambda^{pm}_{nl}e^{-i(\Omega^{lq}_{pq}t_1+\Omega^{pm}_{nl}t_2)}
+ \Lambda^{lm}_{pq}\Lambda^{pq}_{nl}e^{-i(\Omega^{lm}_{pq}t_1+\Omega^{pq}_{nl}t_2)}
\right) + \mathrm{c.c.} \right] \nonumber\\
&&\qquad + \frac{1}{V^4}\sum_l\sum_{n,m,j,k\le K}\left[\Lambda^{ln}_{pm}\Lambda^{pj}_{kl}e^{-i(\Omega^{ln}_{pm}t_1+\Omega^{pj}_{kl}t_2)}{\cal N}_p\alpha^*_m\alpha^*_k\alpha_j \alpha_n + \mathrm{c.c.} \right] \nonumber\\
&&\qquad + \frac{1}{2}\frac{1}{V^4}\sum_l\sum_{n,m,j,k\le K}\left[\Lambda^{ln}_{km}\Lambda^{pj}_{pl}e^{-i(\Omega^{ln}_{km}t_1+\Omega^{pj}_{pl}t_2)}{\cal N}_p\alpha^*_m\alpha^*_k\alpha_j \alpha_n + \mathrm{c.c.} \right] 
 \quad \mathrm{for}\quad p> K. \label{eqA-10}
 \end{eqnarray}
The fourth line of Eq.~\eqref{eqA-4} becomes
\begin{eqnarray}
&&-\frac{1}{4}\frac{1}{V^7}\sum_{nmqjkl}\left[\Lambda^{mn}_{ql}\Lambda^{pl}_{kj}e^{-i(\Omega^{mn}_{ql}t_1+\Omega^{pl}_{kj}t_2)}\langle a_k^{\dagger}a_j^{\dagger}a_q^{\dagger}a_pa_ma_n\rangle+\mathrm{c.c.}\right] \nonumber\\
&&\quad= -\frac{1}{2}\frac{1}{V^4}\sum_{l}\sum_{q\le K}\sum_{k>K}\left[\Lambda^{kk}_{kl}\Lambda^{pl}_{qk}e^{-i(\Omega^{kk}_{kl}t_1+\Omega^{pl}_{qk}t_2)}{\cal N}_k({\cal N}_k-1)\alpha^*_q\alpha_p + \mathrm{c.c.} \right] \nonumber\\
&&\qquad - \frac{1}{4}\frac{1}{V^4}\sum_{l}\sum_{q\le K}\sum_{k>K}\left[\Lambda^{kk}_{ql}\Lambda^{pl}_{kk}e^{-i(\Omega^{kk}_{ql}t_1+\Omega^{pl}_{kk}t_2)}{\cal N}_k({\cal N}_k-1)\alpha^*_q\alpha_p + \mathrm{c.c.} \right] \nonumber\\
&& \qquad -\frac{1}{V^4}\sum_{l}\sum_{q\le K}\sum_{m\ne k,m>K}\sum_{k>K}\left[\Lambda^{km}_{ml}\Lambda^{pl}_{qk}e^{-i(\Omega^{km}_{ml}t_1+\Omega^{pl}_{qk}t_2)}{\cal N}_k{\cal N}_m\alpha_q^*\alpha_p + \mathrm{c.c.} \right] \nonumber\\
&& \qquad -\frac{1}{2}\frac{1}{V^4}\sum_{l}\sum_{q\le K}\sum_{m\ne k,m>K}\sum_{k>K}\left[\Lambda^{km}_{ql}\Lambda^{pl}_{km}e^{-i(\Omega^{km}_{ql}t_1+\Omega^{pl}_{km}t_2)}{\cal N}_k{\cal N}_m\alpha_q^*\alpha_p + \mathrm{c.c.} \right] \nonumber\\
&& \qquad -\frac{1}{V^4}\sum_{l}\sum_{m,n,k\le K}\sum_{q>K}\left[\Lambda^{qk}_{nl}\Lambda^{pl}_{qm}e^{-i(\Omega^{qk}_{nl}t_1+\Omega^{pl}_{qm}t_2)}{\cal N}_q\alpha_m^*\alpha_n^*\alpha_p\alpha_k + \mathrm{c.c.} \right] \nonumber\\
 && \qquad -\frac{1}{2}\frac{1}{V^4}\sum_{l}\sum_{m,n,k\le K}\sum_{q>K}\left[\Lambda^{qk}_{ql}\Lambda^{pl}_{mn}e^{-i(\Omega^{qk}_{ql}t_1+\Omega^{pl}_{mn}t_2)}{\cal N}_q\alpha_m^*\alpha_n^*\alpha_p\alpha_k + \mathrm{c.c.} \right] \nonumber\\
  &&\qquad - \frac{1}{4}\frac{1}{V^4}\sum_{l}\sum_{q,m,k,j,n\le K}\left[\Lambda^{mn}_{ql}\Lambda^{pl}_{kj}e^{-i(\Omega^{mn}_{ql}t_1+\Omega^{pl}_{kj}t_2)}
 \alpha^*_k\alpha^*_j\alpha^*_q\alpha_p\alpha_m\alpha_n + \mathrm{c.c.} \right] \quad \mathrm{for}\quad p\le K, \label{eqA-11}
\end{eqnarray}
or
\begin{eqnarray}
&&-\frac{1}{4}\frac{1}{V^7}\sum_{nmqjkl}\left[\Lambda^{mn}_{ql}\Lambda^{pl}_{kj}e^{-i(\Omega^{mn}_{ql}t_1+\Omega^{pl}_{kj}t_2)}\langle a_k^{\dagger}a_j^{\dagger}a_q^{\dagger}a_pa_ma_n\rangle+\mathrm{c.c.}\right] \nonumber\\
&&\quad= -\frac{1}{4}\frac{1}{V^4}\sum_{l}\left[\Lambda^{lp}_{pp}\Lambda^{pp}_{pl}e^{-i(\Omega^{lp}_{pp}t_1+\Omega^{pp}_{pl}t_2)}{\cal N}_p({\cal N}_p-1)({\cal N}_p-2) + \mathrm{c.c.} \right] \nonumber\\
&&\qquad - \frac{1}{2}\frac{1}{V^4}\sum_{l}\sum_{q\ne p,q> K}\left[\Lambda^{pq}_{ql}\Lambda^{pl}_{pp}e^{-i(\Omega^{pq}_{ql}t_1+\Omega^{pl}_{pp}t_2)}{\cal N}_p({\cal N}_p-1){\cal N}_q + \mathrm{c.c.} \right] \nonumber\\
&&\qquad - \frac{1}{V^4}\sum_{l}\sum_{q\ne p,q> K}\left[\Lambda^{pq}_{pl}\Lambda^{pl}_{qp}e^{-i(\Omega^{pq}_{pl}t_1+\Omega^{pl}_{qp}t_2)}{\cal N}_p({\cal N}_p-1){\cal N}_q + \mathrm{c.c.} \right] \nonumber\\
&&\qquad - \frac{1}{4}\frac{1}{V^4}\sum_{l}\sum_{q\ne p,q>K}\left[\Lambda^{qq}_{pl}\Lambda^{pl}_{qq}e^{-i(\Omega^{qq}_{pl}t_1+\Omega^{pl}_{qq}t_2)}{\cal N}_q({\cal N}_q-1){\cal N}_p + \mathrm{c.c.} \right] \nonumber\\
&&\qquad - \frac{1}{2}\frac{1}{V^4}\sum_{l}\sum_{q\ne p,q>K}\left[\Lambda^{qq}_{ql}\Lambda^{pl}_{pq}e^{-i(\Omega^{qq}_{ql}t_1+\Omega^{pl}_{pq}t_2)}{\cal N}_q({\cal N}_q-1){\cal N}_p + \mathrm{c.c.} \right] \nonumber\\
&&\qquad - \frac{1}{V^4}\sum_{l}\sum_{q\ne p,q>K}\sum_{k\ne q,k\ne p,k>K}\left[\Lambda^{kq}_{kl}\Lambda^{pl}_{pq}e^{-i(\Omega^{kq}_{kl}t_1+\Omega^{pl}_{pq}t_2)}{\cal N}_p{\cal N}_q{\cal N}_k + \mathrm{c.c.} \right] \nonumber\\
&&\qquad - \frac{1}{2}\frac{1}{V^4}\sum_{l}\sum_{q\ne p,q>K}\sum_{k\ne q,k\ne p,k>K}\left[|\Lambda^{qk}_{pl}|^2e^{-i\Omega^{qk}_{pl}(t_1-t_2)}{\cal N}_p{\cal N}_q{\cal N}_k + \mathrm{c.c.} \right] \nonumber
\end{eqnarray}
\begin{eqnarray}
&&\qquad -\frac{1}{V^4}\sum_l\sum_{n,m\le K}\left[\Lambda^{mp}_{pl}\Lambda^{pl}_{np}e^{-i(\Omega^{mp}_{pl}t_1+\Omega^{pl}_{np}t_2)}{\cal N}_p({\cal N}_p-1)\alpha_n^*\alpha_m + \mathrm{c.c.} \right] \nonumber\\
&&\qquad -\frac{1}{2}\frac{1}{V^4}\sum_l\sum_{n,m\le K}\left[\Lambda^{mp}_{nl}\Lambda^{pl}_{pp}e^{-i(\Omega^{mp}_{nl}t_1+\Omega^{pl}_{pp}t_2)}{\cal N}_p({\cal N}_p-1)\alpha_n^*\alpha_m + \mathrm{c.c.} \right] \nonumber\\
&&\qquad -\frac{1}{V^4}\sum_{l}\sum_{n,m\le K}\sum_{q\ne p,q>K}\left[{\cal N}_p{\cal N}_q\alpha_n^*\alpha_m  \right.\nonumber\\
&&\qquad\quad \left.\times\left(\Lambda^{mq}_{ql}\Lambda^{pl}_{np}e^{-i(\Omega^{mq}_{ql}t_1+\Omega^{pl}_{np}t_2)}
+ \Lambda^{mq}_{pl}\Lambda^{pl}_{nq}e^{-i(\Omega^{mq}_{pl}t_1+\Omega^{pl}_{nq}t_2)}
+\Lambda^{mq}_{nl}\Lambda^{pl}_{pq}e^{-i(\Omega^{mq}_{nl}t_1+\Omega^{pl}_{pq}t_2)} 
\right) + \mathrm{c.c.} \right] \nonumber\\
&&\qquad - \frac{1}{2}\frac{1}{V^4}\sum_l\sum_{n,m,j,k\le K}\left[\Lambda^{jn}_{kl}\Lambda^{pl}_{pm}e^{-i(\Omega^{jn}_{kl}t_1+\Omega^{pl}_{pm}t_2)}{\cal N}_p\alpha^*_m\alpha^*_k\alpha_j \alpha_n + \mathrm{c.c.} \right] \nonumber\\
&&\qquad - \frac{1}{4}\frac{1}{V^4}\sum_l\sum_{n,m,j,k\le K}\left[\Lambda^{jn}_{pl}\Lambda^{pl}_{mk}e^{-i(\Omega^{jn}_{pl}t_1+\Omega^{pl}_{mk}t_2)}{\cal N}_p\alpha^*_m\alpha^*_k\alpha_j \alpha_n + \mathrm{c.c.} \right] 
 \quad \mathrm{for}\quad p> K. \label{eqA-12}
 \end{eqnarray}
Finally, the fifth line of Eq.~\eqref{eqA-4} becomes 
 \begin{eqnarray}
&&-\frac{1}{4}\frac{1}{V^7}\sum_{nmqjkl}\left[\Lambda^{mn}_{pq}\Lambda^{pj}_{kl}e^{-i(\Omega^{mn}_{pq}t_1+\Omega^{pj}_{kl}t_2)}\langle a_k^{\dagger}a_l^{\dagger}a_q^{\dagger}a_ja_ma_n\rangle+\mathrm{c.c.}\right] \nonumber\\
&&\quad= -\frac{1}{4}\frac{1}{V^4}\sum_{q>K}\left[\Lambda^{qq}_{pq}\Lambda^{pq}_{qq}e^{-i(\Omega^{qq}_{pq}t_1+\Omega^{pq}_{qq}t_2)}{\cal N}_q({\cal N}_q-1)({\cal N}_q-2) + \mathrm{c.c.} \right] \nonumber\\
&&\qquad - \frac{1}{V^4}\sum_{q>K}\sum_{k\ne q,k> K}\left[\Lambda^{qk}_{qp}\Lambda^{pq}_{kq}e^{-i(\Omega^{qk}_{qp}t_1+\Omega^{pq}_{kq}t_2)}{\cal N}_q({\cal N}_q-1){\cal N}_k + \mathrm{c.c.} \right] \nonumber\\
&&\qquad - \frac{1}{2}\frac{1}{V^4}\sum_{q>K}\sum_{k\ne q,k> K}\left[\Lambda^{qk}_{pk}\Lambda^{pq}_{qq}e^{-i(\Omega^{qk}_{pk}t_1+\Omega^{pq}_{qq}t_2)}{\cal N}_q({\cal N}_q-1){\cal N}_k + \mathrm{c.c.} \right] \nonumber\\
&&\qquad - \frac{1}{2}\frac{1}{V^4}\sum_{q>K}\sum_{k\ne q,k>K}\left[\Lambda^{qq}_{pq}\Lambda^{pk}_{kq}e^{-i(\Omega^{qq}_{pq}t_1+\Omega^{pk}_{kq}t_2)}{\cal N}_q({\cal N}_q-1){\cal N}_k + \mathrm{c.c.} \right] \nonumber\\
&&\qquad - \frac{1}{4}\frac{1}{V^4}\sum_{q>K}\sum_{k\ne q,k>K}\left[\Lambda^{qq}_{pk}\Lambda^{pk}_{qq}e^{-i(\Omega^{qq}_{pk}t_1+\Omega^{pk}_{qq}t_2)}{\cal N}_q({\cal N}_q-1){\cal N}_k + \mathrm{c.c.} \right] \nonumber\\
&&\qquad - \frac{1}{V^4}\sum_{q>K}\sum_{k\ne q,k>K}\sum_{m\ne k,m\ne q,m>K}\left[\Lambda^{km}_{pm}\Lambda^{pq}_{qk}e^{-i(\Omega^{km}_{pm}t_1+\Omega^{pq}_{qk}t_2)}{\cal N}_q{\cal N}_k{\cal N}_m + \mathrm{c.c.} \right] \nonumber\\
&&\qquad - \frac{1}{2}\frac{1}{V^4}\sum_{q>K}\sum_{k\ne q,k>K}\sum_{m\ne k,m\ne q,m>K}\left[|\Lambda^{km}_{pq}|^2e^{-i\Omega^{km}_{pq}(t_1-t_2)}{\cal N}_q{\cal N}_k{\cal N}_m + \mathrm{c.c.} \right] \nonumber\\
&&\qquad -\frac{1}{V^4}\sum_{q>K}\sum_{n,m\le K}\left[\Lambda^{mq}_{pq}\Lambda^{pq}_{nq}e^{-i(\Omega^{mq}_{pq}t_1+\Omega^{pq}_{nq}t_2)}{\cal N}_q({\cal N}_q-1)\alpha_n^*\alpha_m + \mathrm{c.c.} \right] \nonumber\\
&&\qquad -\frac{1}{2}\frac{1}{V^4}\sum_{q>K}\sum_{n,m\le K}\left[{\cal N}_q({\cal N}_q-1)\alpha_n^*\alpha_m
\left(\Lambda^{qq}_{pq}\Lambda^{pm}_{nq}e^{-i(\Omega^{qq}_{pq}t_1+\Omega^{pm}_{nq}t_2)} + \Lambda^{mq}_{pn}\Lambda^{pq}_{qq}e^{-i(\Omega^{mq}_{pn}t_1+\Omega^{pq}_{qq}t_2)}\right) + \mathrm{c.c.} \right] \nonumber\\
&&\qquad -\frac{1}{4}\frac{1}{V^4}\sum_{q>K}\sum_{n,m\le K}\left[\Lambda^{qq}_{pn}\Lambda^{pm}_{qq}e^{-i(\Omega^{qq}_{pn}t_1+\Omega^{pm}_{qq}t_2)}{\cal N}_q({\cal N}_q-1)\alpha_n^*\alpha_m + \mathrm{c.c.} \right] \nonumber\\
&&\qquad -\frac{1}{V^4}\sum_{q>K}\sum_{n,m\le K}\sum_{k\ne q,k>K}\left[{\cal N}_q{\cal N}_k\alpha_n^*\alpha_m \left(\Lambda^{qk}_{pk}\Lambda^{pm}_{nq}e^{-i(\Omega^{qk}_{pk}t_1+\Omega^{pm}_{nq}t_2)} \right. \right.\nonumber\\
&&\qquad\qquad \left. \left.+ \Lambda^{mk}_{pk}\Lambda^{pq}_{nq}e^{-i(\Omega^{mk}_{pk}t_1+\Omega^{pq}_{nq}t_2)}
+ \Lambda^{mk}_{pq}\Lambda^{pq}_{nk}e^{-i(\Omega^{mk}_{pq}t_1+\Omega^{pq}_{nk}t_2)}
+\Lambda^{mk}_{pn}\Lambda^{pq}_{qk}e^{-i(\Omega^{mk}_{pn}t_1+\Omega^{pq}_{qk}t_2)} 
\right) + \mathrm{c.c.} \right] \nonumber\\
&&\qquad -\frac{1}{2}\frac{1}{V^4}\sum_{q>K}\sum_{n,m\le K}\left[\Lambda^{qk}_{pn}\Lambda^{pm}_{qk}e^{-i(\Omega^{qk}_{pn}t_1+\Omega^{pm}_{qk}t_2)}{\cal N}_q{\cal N}_k\alpha_n^*\alpha_m + \mathrm{c.c.} \right] \nonumber\\
&&\qquad - \frac{1}{V^4}\sum_{q>K}\sum_{k,l,m,n\le K}\left[\Lambda^{qn}_{pl}\Lambda^{pm}_{qk}e^{-i(\Omega^{qn}_{pl}t_1+\Omega^{pm}_{qk}t_2)}{\cal N}_q\alpha^*_k\alpha^*_l\alpha_m \alpha_n + \mathrm{c.c.} \right] \nonumber
\end{eqnarray}
\begin{eqnarray}
&&\qquad - \frac{1}{2}\frac{1}{V^4}\sum_{q>K}\sum_{k,l,m,n\le K}\left[{\cal N}_q\alpha^*_k\alpha^*_l\alpha_m \alpha_n
\left(\Lambda^{mn}_{pl}\Lambda^{pq}_{qk}e^{-i(\Omega^{mn}_{pl}t_1+\Omega^{pq}_{qk}t_2)} + \Lambda^{qn}_{pq}\Lambda^{pm}_{kl}e^{-i(\Omega^{qn}_{pq}t_1+\Omega^{pm}_{kl}t_2)} \right) + \mathrm{c.c.} \right] \nonumber\\
&&\qquad - \frac{1}{4}\frac{1}{V^4}\sum_{q>K}\sum_{k,l,m,n\le K}\left[\Lambda^{mn}_{pq}\Lambda^{pq}_{kl}e^{-i(\Omega^{mn}_{pq}t_1+\Omega^{pq}_{kl}t_2)}{\cal N}_q\alpha^*_k\alpha^*_l\alpha_m\alpha_n + \mathrm{c.c.} \right] \nonumber\\
&&\qquad - \frac{1}{4}\frac{1}{V^4}\sum_{k,l,q,j,m,n\le K}\left[\Lambda^{mn}_{pq}\Lambda^{pj}_{kl}e^{-i(\Omega^{mn}_{pq}t_1+\Omega^{pj}_{kl}t_2)}\alpha^*_k\alpha^*_l\alpha^*_q\alpha_j\alpha_m\alpha_n + \mathrm{c.c.} \right].
\label{eqA-13}
\end{eqnarray}

In the particle kinetic regime, these tremendously long equations can be simplified as follows.
For $p>K$, we obtain
\begin{eqnarray}
{\cal N}_p(t) &\simeq& {\cal N}_p(t_0)  - \int^t_{t_0}dt_2\int^{t_2}_{t_0}dt_1\langle[H_I(t_1),[H_I(t_2),{\cal \hat{N}}_p]]\rangle. \label{eqA-14}
\end{eqnarray}
In general, terms which contribute to the expectation value take a form
\begin{equation}
\langle[H_I(t_1),[H_I(t_2),{\cal \hat{N}}_p]]\rangle \propto e^{-i(\Omega_1t_1+\Omega_2 t_2)} + \mathrm{c.c.} \nonumber
\end{equation}
Taking the time derivative after performing the integration over $t_1$ and $t_2$, we find
\begin{equation}
\frac{d{\cal N}_p}{dt} \propto \frac{i}{\Omega_1}e^{-i(\Omega_1+\Omega_2)t}-\frac{i}{\Omega_1}e^{-i(\Omega_1t_0+\Omega_2t)}+\mathrm{c.c.} \label{eqA-15}
\end{equation}
If $\Omega_1+\Omega_2\ne 0$, these terms drop out because of the rapidly oscillating factor in the particle kinetic regime ($\Omega^{kl}_{pq}t\to \infty$).
On the other hand, if $\Omega_1+\Omega_2 = 0$, the first term of the right-hand side of Eq.~\eqref{eqA-15} cancels with its complex conjugate.
Then we obtain
\begin{equation}
\frac{d{\cal N}_p}{dt} \propto \frac{2}{\Omega_1}\sin\Omega_1(t-t_0). \nonumber
\end{equation}
Note that the energy conservation emerges in the limit because
$\Omega_1(t-t_0)\to\infty$,
\begin{equation}
\frac{2}{\Omega_1}\sin \Omega_1(t-t_0) \to 2\pi\delta(\Omega_1), \label{eqA-16}
\end{equation}
which implies that terms with $\Omega_1\ne 0$ do not contribute to the
final result in this limit.  For example, the second line of
Eq.~\eqref{eqA-7} gives $\Omega_1=\Omega^{qm}_{pp}$, which does not
vanish because of the conservation law of three momenta in
$\Lambda^{qm}_{pp}$. The exception is the case with $q=m=p$, but the
careful inspection shows that this term exactly cancels with the second
line of Eq.~\eqref{eqA-8}. Similar discussions are applied for the
second, third, fourth, and fifth lines of Eq.~\eqref{eqA-10}, the second
and fourth lines of Eq.~\eqref{eqA-12}, and the second, third, and sixth
lines of Eq.~\eqref{eqA-13}. After all, the remaining terms lead to
\begin{equation}
\frac{d{\cal N}_p}{dt} = \frac{1}{2V^4}\sum_{klq>K}|\Lambda^{kl}_{pq}|^22\pi\delta(\Omega^{kl}_{pq})\left[{\cal N}_k{\cal N}_l({\cal N}_p+1)({\cal N}_q+1) - ({\cal N}_k+1)({\cal N}_l+1){\cal N}_p{\cal N}_q \right], \label{eqA-17}
\end{equation}
where we have neglected the contribution that contains the integration
over condensed modes (i.e. $\sum_{q\le K}\sum_{k,l>K}$), because such
terms are prohibited by the energy conservation~\eqref{eqA-16}.  In this
way, we recover the usual Boltzmann equation~\cite{Erken:2011dz}.


\end{document}